\documentclass[11pt,a4paper]{article}
\pdfoutput=1
\usepackage{multirow}
\usepackage{jheppub}
\usepackage{enumerate}
\usepackage{relsize}
\usepackage{mathrsfs}
\usepackage{inputenc}
\usepackage{braket}
\usepackage{graphicx}
\usepackage{chemarrow}
\usepackage{mathtools}
\usepackage{mathrsfs}
\usepackage{amsfonts}
\usepackage{caption}
\usepackage{blindtext}
\usepackage{bbold,amsmath,lipsum,amssymb,leftindex}
\usepackage[T1]{fontenc}
\usepackage{comment}
\usepackage{rotating}
\usepackage{tikz}
\usepackage{capt-of}
\usepackage{bm}
\usepackage{float}
\usepackage{relsize}
\usepackage{array}
\usepackage{amsfonts}
\usepackage{amsmath}
\usepackage{setspace}
\usepackage{easybmat}
\usepackage{slashed}
\usepackage{amsmath}
\usepackage{amssymb}
\usepackage{graphicx}
\usepackage{verbatim}
\usepackage{caption}
\usepackage{subcaption}
\usepackage{makecell}

\usetikzlibrary{decorations.pathmorphing}
\usetikzlibrary{positioning}
\usetikzlibrary{intersections}

\def\spa#1{\phantom{\fbox{\rule[-#1cm]{0cm}{0cm}}}}

\newcommand{\beqa}{\begin{eqnarray}}
\newcommand{\eeqa}{\end{eqnarray}}
\def\f{\frac}

\title{Scalar fields and 3D Flat Space Cosmologies}
\author[a]{Arjun Bagchi,} \author[a]{Supratik Biswas,}\author[a]{Astha Kakkar,}  \author[a,b,c]{and Saikat Mondal.} \author{\\}
\affiliation[a]{Indian Institute of Technology Kanpur, Kalyanpur, Kanpur 208016, INDIA.}
\affiliation[b]{Erwin Schrödinger International Institute for Mathematics and Physics, University of Vienna, 1090 Vienna, AUSTRIA.}
\affiliation[c]{Institute for Theoretical Physics, TU Wien, Wiedner Hauptstrasse 8–10/136, A-1040 Vienna,
AUSTRIA.}

\emailAdd{abagchi@iitk.ac.in, supratikb22@iitk.ac.in, asthakakkar8@gmail.com, saikatmd@iitk.ac.in}

\abstract{Flat Space Cosmologies (FSC) are time-dependent solutions in Einstein gravity in three-dimensional (3D) spacetimes with zero cosmological constant. These are orbifolds of 3D flat space that have a cosmological horizon and can be thought of as analogs of the Banados-Tietelboim-Zanelli (BTZ) black holes of AdS$_3$. We study scalar perturbations about these FSC solutions and explore the spectrum of quasi-normal modes (QNMs) crucially treating the cosmological horizon as a hard wall and extending to complex momenta. We connect this intrinsic analysis with the flatspace limit of the corresponding analysis in the BTZ black hole. The FSC QNMs are then utilized to build the scalar one-loop partition function by methods pioneered by Denef, Hartnoll and Sachdev in various simplifying limits and compared with existing answers in the literature.}

\preprint{}

\begin{document}
\maketitle

\section{Introduction}
All astrophysical objects in the universe oscillate in some specific way. Observations of such oscillations for different stars encode rich information about the internal structure of the star. In the case of very dense stellar objects such as neutron stars and black holes, these oscillations can be detected by gravitational wave detectors. Recent discoveries by LIGO and VIRGO \cite{PhysRevLett.116.061102} have led to a revolution in this field. In addition, when two black holes merge, they generate gravitational waves that carry information about the properties of the black holes, viz., mass, spin, etc. After this merger, the newly formed black hole undergoes a ringdown phase post which it settles into a stable state. This phase is characterized by oscillations that can be described by what are known as quasi-normal modes (QNMs). 

\medskip

Within the context of general relativity, QNMs are known to emerge as perturbations of black hole spacetime. It is well known that when a black hole is perturbed, the geometry around it undergoes damped oscillations. The frequencies and damping times of these oscillations do not depend on the initial perturbations and can be determined entirely by the black hole parameters. These oscillations give rise to the QNMs. The main trait that sets them apart from normal modes is that when the system is open, energy loss can occur through the emission of gravitational waves. Because of this, the frequencies become complex, with the real part standing as the actual frequency of the oscillation while the imaginary part represents the damping. 

\medskip

The research on black hole perturbations was initiated by Regge and Wheeler \cite{PhysRev.108.1063} and then carried forward by Zerilli \cite{Zerilli:1970se, Zerilli:1970wzz, Zerilli:1974ai} and Chandrasekhar \cite{Chandrasekhar:1985kt} to name a few. The first mention of QNMs came in the case of black holes in asymptotically flat spacetimes \cite{Vishveshwara:1970zz, Press:1971wr}. Since then, there has been an abundance of research on the QNMs of black holes in both asymptotically Anti-de Sitter (AdS) as well as asymptotically de-Sitter (dS) spacetimes. Interested readers are directed for a review on this topic to \cite{Kokkotas:1999bd} and for a recent work to \cite{Berti:2009kk}.  

\medskip

Other than the astrophysical interest of studying QNMs, from the theoretical viewpoint, it has found its usefulness in the understanding of quantum gravity, particularly via the holographic principle \cite{tHooft:1993Dmi, Susskind:1994vu} and the AdS/CFT correspondence \cite{Maldacena:1997re}.
A duality between QNMs of AdS and the timescale of approach to thermal equilibrium in the dual CFT was first proposed numerically in \cite{Horowitz:1999jd}. In the context of AdS$_3$/CFT$_2$, it was shown that the QNMs of the Banados-Teitleboim-Zanelli (BTZ) black holes \cite{Banados_1992, Banados_1993} match with the position of the poles of the retarded Green's function, which describe the linear response in 2D CFT \cite{Birmingham:2001pj}.

\medskip

There has been a lot of recent interest in understanding a version of holography that would be of relevance for asymptotically flat spacetimes (AFS) and progress has principally been achieved in the so called Celestial and Carrollian approaches. The Celestial approach puts forward a co-dimension two dual relativistic CFT as a potential dual \cite{Strominger:2013jfa,He:2014laa,Strominger:2014pwa,Strominger:2017zoo, Pasterski:2021rjz}, whereas a co-dimension one Carrollian CFT (CCFT) is the other candidate \cite{Bagchi:2016bcd}. While as duals to 4d AFS, the Celestial approach has been more popular, resulting in a lot of new insights about asymptotic symmetries and scattering amplitudes, there has been a recent resurgence in work in the Carrollian approach following \cite{Bagchi:2022emh, Donnay:2022aba}. An incomplete list of references include \cite{Bagchi:2023fbj,Donnay:2022wvx,Mason:2023mti,Saha:2023hsl,Nguyen:2023vfz, Bagchi:2023cen}. For a more complete list of references and further details, the reader is referred to the recent review \cite{Bagchi:2025vri}.

\medskip

Carrollian holography, which predates the Celestial programme and began by the observation of the isomorphism of asymptotic symmetry algebras of flat spacetimes viz. Bondi-van der Burgh-Metzner-Sachs (BMS) algebras \cite{Bondi:1962px, Sachs:2} with lower dimensional non-Lorentzian algebras \cite{Bagchi:2010eg, Bagchi:2012cy, Duval_2014}, had early success in the context of 3D AFS where various important checks like the matching between bulk quantities and boundary ones have been achieved for thermal entropy \cite{Bagchi:2012xr}\footnote{See also \cite{Barnich:2012xq}.}, entanglement entropy \cite{Bagchi:2014iea,Jiang:2017ecm}, stress-tensor correlations \cite{Bagchi:2015wna}. Some other important advances include \cite{Barnich:2012aw, Bagchi:2012yk, Afshar:2013vka, Gonzalez:2013oaa, Barnich:2015mui, Hartong:2015usd}. In this paper, we would be concerned with bulk analysis in 3D AFS and make some remarks about 2D CCFT in our conclusions. 

\medskip

As mentioned above, the BTZ black holes are the typical objects of interest in the AdS$_3$/CFT$_2$ correspondence. This is especially true since 3D Einstein gravity does not contain any propagating degrees of freedom. The analogs of the BTZ black holes for 3D AFS are the so-called {\em Flat Space Cosmology} (FSC) solutions \cite{Cornalba_2002, Cornalba_2004}. Like BTZ, these can be thought of as orbifolds of 3D Minkowski spacetimes. Just as Minkowski spacetime is obtained by an infinite radius limit of AdS, FSCs can be obtained in the same limit from the (non-extremal) BTZ solution. These FSC solutions are endowed with a cosmological horizon and hence can be attributed with thermodynamical quantities and specifically an entropy, which can be readily derived from a Cardy-like counting of states in the dual CCFT \cite{Bagchi:2012xr}. 

\medskip

In this paper, we will be interested in scalar perturbations of these FSC solutions and will investigate quasi-normal modes associated with such perturbations. Previous attempts at constructing QNM for the FSC have taken indirect routes, e.g. understanding torus one-point and two-point functions in the dual theory \cite{Bagchi:2020rwb, Bagchi:2023uqm} and using the Selberg zeta function \cite{Bagchi:2023ilg}. Here we adopt a more direct approach and attempt to solve the problem in the bulk 3D AFS by putting appropriate boundary conditions on the cosmological horizon. This apparently straight-forward route was previously unsuccessful as it was unclear what boundary conditions would be appropriate at the cosmological horizon. As we show later in the paper, for the obvious choices of purely ingoing or purely outgoing boundary conditions, we get trivial solutions. In this paper, we use a hard wall at the cosmological horizon. With this, and complexified momenta, we would be able to generate an interesting set of QNMs for the FSC. 

\begin{table}[t]
    \centering
     \setlength{\leftmargini}{0.4cm}
    \begin{tabular}{| m{2.25cm} | m{3.95cm} | m{3.4cm} | m{3.6cm} |}
        \hline
       \textbf{Features} & \textbf{~~~~~~~~~~~~BTZ} & \textbf{~~~~~~~~dS} & \textbf{~~~~~~~FSC} \\
        \hline
        \textbf{Boundary conditions} & 
        \begin{enumerate}
            \item Ingoing at event $~~~~~~$ horizon 
            \item Reflective at infinity 
        \end{enumerate} & 
        \begin{enumerate} 
            \item Outgoing at cosmological horizon  
            \item Regular at origin            
        \end{enumerate} & 
        \begin{enumerate} 
          \item Reflective at cosmological horizon 
         \item Bounded at infinity
        \end{enumerate} \\[10pt]
        \hline
        \textbf{Solutions} & \thead{$\psi=e^{i\omega t-k\phi}z^{-\frac{ik_+\ell}{2}}$\\$(1-z)^{1-\frac{\Delta}{2}}{}_2F_1(a,b;c;z)$}  & \thead{$\psi=e^{-i\omega t}Y_{l,m}y^{\f{l}{2}}$\\$(1-y)^{\f{i\omega\ell}{2}}~_2F_1(a,b,c;y) $} & \thead{$   \psi=J_{\pm\nu}(\omega|\tau|)e^{ipx}e^{i\frac{n}{r_0}y}$}\\[10pt]
\hline
\textbf{Quasi-Normal Modes} & \thead{$\omega_{p,s,\pm}=\pm \f{p}{L}-2\pi Ti(\Delta+2s)$\\ \\ $s=0,1,2\cdots, $\\
$p=0,\pm1,\pm2,\cdots$ }  & \thead{$\omega_{n,l_{\pm}}=-i\f{2n+l+\Delta}{L}$ \\ \\ $n,l=0,1,\cdots$} & \thead{$\omega_{i,r}^2=\sqrt{A_{n,N}^2+B_{n,N}^2}$ \\ $\pm A_{n,N}$ \\ \\ $N,n=0,\pm 1,\cdots$ }\\[10pt]
\hline
\textbf{Details in} & Appendix \ref{APPA} & Appendix \ref{dsqnm} & Section \ref{qnmcalc}\\[10pt]
\hline
    \end{tabular}
    \caption{Comparison between BTZ, dS and FSC }
\label{compare}
\end{table}

Table \ref{compare} provides a comparative analysis of perturbations of BTZ black holes, de-Sitter spacetimes (with cosmological horizons), and our object of interest in the paper, FSC. Here we summarize how the different geometries require different boundary conditions to compute the corresponding QNMs. FSCs, as we shall see, have features in common with both BTZ and dS, but require a different set of boundary conditions that respect the physics of their geometry. For the notations and detailed analysis please refer to appendix \ref{APPA} and \ref{dsqnm} for BTZ and dS, respectively, and section \ref{scalar} for FSC.

\medskip

Once equipped with the QNMs of these FSCs and having established the regularity of our solutions, we build the one-loop partition functions using the Denef-Hartnoll-Sachdev \cite{Denef:2009kn} prescription. We find closed-form answers in certain regimes of parameter space and connect to existing results in the literature \cite{Barnich:2015mui, Bagchi:2023ilg}. 

\medskip

The outline of the paper is as follows. In section \ref{orbifold}, we provide a brief review of FSCs and describe how these FSCs arise as shifted-boost orbifolds of Minkowski spacetime. In section \ref{scalar}, we study the scalar field solutions in FSC background. Section \ref{fscwf} discusses the scalar wavefunctions, \ref{rbc} imposes the required boundary conditions and we find the quasi-normal modes in \ref{qnmcalc}. Section \ref{limit} provides the limiting analysis of the results from the BTZ black hole analysis. In section \ref{dhs}, we build the scalar one-loop partition function in FSC using the quasi-normal modes and the DHS method. We then conclude in section \ref{conc}. The justification for our choice of boundary conditions is given in appendix \ref{inout}. A brief analysis of the scalar field in region II of the FSC geometry is provided in appendix \ref{innerfsc}. Reviews of scalar field solutions and one-loop partition functions using the DHS method in AdS, BTZ, and dS backgrounds are provided in appendix \ref{adsqnm}, \ref{APPA},  and \ref{dsqnm} respectively.

\section{3D Flatspace and Cosmological Solutions}\label{orbifold} 
Despite the absence of propagating degrees of freedom, 3D Einstein gravity is a very useful playground for building our understanding of quantum aspects of gravity in more relevant higher dimensions. We begin with a brief review of 3D gravity in asymptotically flat spacetimes to set up notation for the rest of the paper. 

\subsection{Anti-de Sitter preliminaries}
The 3D Einstein-Hilbert action with a cosmological constant is
\begin{equation}
    \mathcal{S}[g] = \frac{1}{16\pi G}\int\,d^3x\sqrt{-g}(R + 2\Lambda) + \mathcal{S}_{bdy}
\end{equation}
where $R$ is Ricci scalar curvature and $\Lambda$ is the cosmological constant. For $\Lambda<0$, the solutions of Einstein's equations are locally AdS. The vacuum solutions are global AdS$_3$ spacetime characterized by the metric
\begin{equation}
    ds^2 = -\bigg(1+\frac{r^2}{\ell^2}\bigg)dt^2 + \bigg(1+\frac{r^2}{\ell^2}\bigg)^{-1}dr^2 + r^2d\phi^2.
\end{equation}
Here $\ell$ is the AdS radius with $\Lambda = -1/\ell^2$. The BTZ black holes, in the non-extremal case, are given by 
\begin{equation}
    ds^2=-\frac{(r^2-r_+^2)(r^2-r_-^2)}{r^2\ell^2}dt^2+\frac{r^2\ell^2dr^2}{(r^2-r_+^2)(r^2-r_-^2)} +r^2\left(d\phi-\frac{r_+r_-}{\ell r^2}dt\right)^2.
\end{equation}
Here $r_{\pm}$ are the outer and inner horizons respectively and are given in terms of mass $M$ and angular momentum $J$ as
\begin{align}
    r_\pm = \sqrt{2G \ell (\ell M + J)} \pm \sqrt{2G \ell(\ell M - J)}.   
\end{align}

\subsection{Flatspace in 3D: Asymptotic symmetries and FSC}
In this paper, we are interested in exploring Einstein's gravity in 3D without a cosmological constant. It goes without saying that we have the global Minkowski spacetime as a solution. Interestingly, along the lines parallel to our discussion above, in addition to this, there also exists another solution which satisfies Einstein's equation given by
\begin{equation}\label{fsc}
    ds^2 = \frac{\hat{r}_+^2(r^2-r_0^2)}{r^2} dt^2 - \frac{r^2 dr^2}{\hat{r}_+^2(r^2-r_0^2)} + r^2\left(d\phi - \frac{r_0\hat{r}_+}{r^2} dt\right)^2. 
\end{equation}
These are cosmological spacetimes with horizons at $r=r_0$ called \textit{flat-space cosmologies} (FSCs) \cite{Cornalba_2002, Cornalba_2004}. There is a singularity at $r=0$, which is timelike. The two parameters $\hat{r}_+$ and $r_0$ are related to the mass $M$ and angular momentum $J$:
\begin{eqnarray}
    \hat{r}_+ = \sqrt{8GM}\,,\qquad r_0 = \sqrt{\frac{2G}{M}}\,J\,.
\end{eqnarray}

Flat space cosmologies form the most general zero mode solutions to 3D AFS with Barnich-Compère boundary conditions \cite{Barnich:2006av}. The generic solution for such boundary conditions is given by a metric of the form 
\begin{align}\label{BC-metric}
    ds^2 = \Theta(\phi) du^2 - 2 du dr + \left( \Xi(\phi) + u \partial_\phi \Theta (\phi)\right) du d\phi + r^2 d\phi^2, 
\end{align}
where $\Theta(\phi)$ is called the mass aspect ratio and $\Sigma(\phi)$ the angular momentum aspect ratio. For the FSC, these are 
\begin{align}
    \Theta(\phi) = M, \quad \Sigma(\phi) = J.
\end{align}
With the above-quoted boundary conditions, a canonical analysis at the null boundary $\mathscr{I}^\pm$ (which has a structure $R_u \times S^1$, with $R_u$ representing a null line) of 3D AFS yields an asymptotic symmetry algebra that is known as the 3D Bondi-van der Burgh-Metzner-Sachs (BMS$_3$) algebra \cite{Barnich:2006av}:
\begin{subequations}
    \begin{align}
    [L_n, L_m] &= (n-m)L_{n+m} + \frac{c_L}{12} \delta_{n+m,0}(n^3 -n), \\
    [L_n, M_m] &= (n-m)M_{m+m} + \frac{c_M}{12} \delta_{n+m,0}(n^3 -n), \quad [M_n, M_m] =0.
\end{align} 
\end{subequations}
Here $L_n$'s are called superrotations and represent the generators of diffeomorphisms of the circle at $\mathscr{I}^\pm$ while $M_n$'s are generators of angle-dependent translations of the null direction and are called supertranslations. $c_L, c_M$ are two allowed central charges associated with the asymptotic symmetry algebra. For the case of Einstein gravity, these turn out to be 
\begin{align}
  c_L=0, \quad c_M = \frac{3}{G},  
\end{align}
where $G$ is the Newton's constant. 

As is very well known from the seminal analysis of Brown and Henneaux \cite{Brown:1986nw}, , the asymptotic symmetry algebra of 3D AdS spacetimes is given by two copies of the Virasoro algebra
\begin{subequations}
 \begin{align}
    [\mathcal{L}_n, \mathcal{L}_n] &= (n-m) \mathcal{L}_{n+m} + \frac{c}{12} \delta_{n+m,0}(n^3 -n), \\
    [\bar{\mathcal{L}}_n, \bar{\mathcal{L}}_m] &= (n-m) \bar{\mathcal{L}}_{n+m} + \frac{\bar{c}}{12} \delta_{n+m,0}(n^3 -n), \quad [{\mathcal{L}}_n, \bar{\mathcal{L}}_m] =0. 
\end{align}   
\end{subequations}
In the above, the central terms for 3D Einstein gravity in AdS$_3$ turn out to be 
\begin{align}
   c = \bar{c} = \frac{3\ell}{2G} 
\end{align}
where $\ell$ is the radius of AdS and $G$ is the Newton's constant.  

One can arrive at 3D AFS from AdS$_3$ by taking an infinite radius limit. This can be understood also in terms of the asymptotic symmetry algebras. One needs to scale the linear combinations of the Virasoro algebras in order to arrive at the BMS algebra:
\begin{align}
    L_n = \mathcal{L}_n - \bar{\mathcal{L}}_{-n}, \quad M_n = \frac{1}{\ell}\left( \mathcal{L}_n + \bar{\mathcal{L}}_{-n} \right), \quad \ell \to \infty. 
\end{align}
It is instructive to check that the central terms also map properly:
\begin{align}
  c_L=  c - \bar{c} = 0, \quad c_M = \frac{1}{\ell}\left( c + \bar{c} \right) = \frac{3}{G}.  
\end{align}

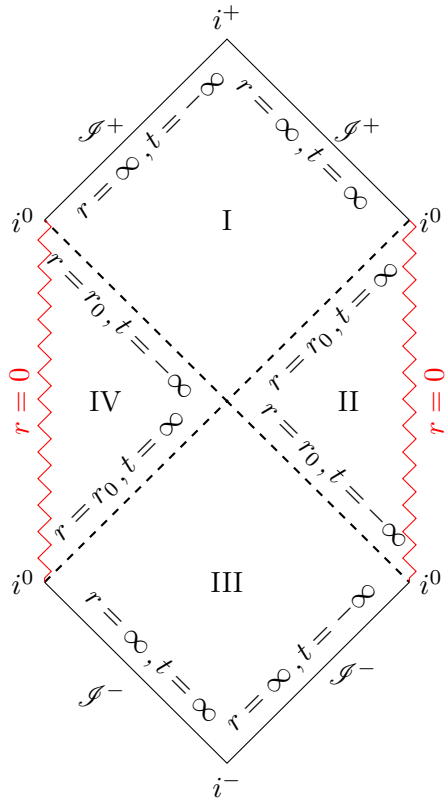
\begin{figure}[t] 
\centering
\begin{tikzpicture}[scale=0.8]
\node (I)    at (0,3)   {I};
\node (II)   at (0,-3)   {III};
\node (III)  at (2,0) {II};
\node (IV)   at (-2,0) {IV};

       ;
\path  
  (II) +(90:3)  coordinate                  (IItop)
       +(-90:3) coordinate[label=-90: $i^-$ ] (IIbot)
       +(0:3)   coordinate[label=360:$i^0$]  (IIright)
       +(180:3) coordinate[label=180:$i^0$] (IIleft)
       ;
\draw 
      (IIright) -- 
          node[midway, above, sloped] {$r=\infty\,,t=-\infty$}
          node[midway, below, right] {$\mathscr{I}^-$}
      (IIbot) --
          node[midway, above, sloped] {$r=\infty\,,t=\infty$}
          node[midway, below left]    {$\mathscr{I}^-$}    
      (IIleft) ;

\draw[thick,dashed]      
       (IIleft) -- 
          node[midway, above, sloped]  {$r=r_0\,,t=\infty$}
      (IItop) --      
       (IIright)  
          node[midway, above, sloped] {$r=r_0\,,t=-\infty$};
\path
   (I) +(90:3)  coordinate [label=90: $i^+$ ] (Itop)
       +(-90:3) coordinate (Ibot)
       +(180:3) coordinate [label=180:$i^0$] (Ileft)
       +(0:3)   coordinate [label=360:$i^0$] (Iright)
       ;
\draw  
    (Ileft) -- 
        node[midway, below, sloped] {$r=\infty\,,t=-\infty$}
          node[midway, above, left] {$\mathscr{I}^+$}
    (Itop) -- 
    (Iright)
        node[midway, below, sloped] {$r=\infty\,,t=\infty$}
          node[midway, above, right] {$\, \mathscr{I}^+$};
\draw[thick,dashed] 
    (Iright) --
        node[midway, below, sloped]  {$r=r_0\,,t=\infty$}
    (Ibot) --  
    (Ileft) 
        node[midway, below, sloped]  {$r=r_0\,,t=-\infty$}; 
\draw[red,decorate,decoration=zigzag] (Ileft) -- (IIleft)
      node[midway, above, inner sep=2mm,left] {\begin{turn}{90}
          $r=0$
      \end{turn}}
    ;
\draw[red,decorate,decoration=zigzag] (Iright) -- (IIright)
      node[midway, below, inner sep=2mm,right] {\begin{turn}{90}
          $r=0$
      \end{turn}};
\end{tikzpicture}
\captionof{figure}{Penrose diagram of FSC solution. Cosmological horizons are depicted as dashed lines. Red squiggly lines and solid black lines correspond to timeline singularities and past-future null infinities.}
\label{pdfsc}
\end{figure}

Following the same chain of logic, one expects the general zero mode solutions of AdS$_3$ to map to those of 3D AFS and indeed FSCs arise as the $\ell\to\infty$ limit of non-extremal BTZ solutions \cite{Cornalba_2002, Barnich:2012aw, Bagchi:2012xr}. In this limit, $r_+ \rightarrow \ell \hat{r}_+$ and $r_- \rightarrow r_0$, implying that the outer horizon is pushed to infinity while the inner horizon stays at a finite distance. The inside of the parent black hole becomes the whole of the spacetime and hence time and radial directions flip around to make this a time-dependent solution in asymptotically flat spacetime and the remnant of the inner horizon becomes a cosmological horizon. The Penrose diagram of FSC is portrayed in figure \ref{pdfsc}.

\subsection{FSC as shifted-boost orbifolds} 
It is well known that BTZ black hole are quotients of AdS$_3$ \cite{Banados_1993}. Similarly, it is also possible to find a locally well-defined transformation that maps between FSCs and the Minkowski spacetime. Since this mapping is a combination of translation (shift) and boost of the coordinates, FSCs are known as shifted-boost orbifolds of Minkowski spacetime \cite{Cornalba_2002}. Below we briefly discuss technical aspects of this construction. A lot of the material including notation and conventions follow \cite{Cornalba_2002}.

$\text{AdS}_3$ can be visualized as an embedding in a 4-dimensional manifold $\mathbb{R}^{(2,2)}$ with the embedding equation $-u^2 - v^2 + x^2 + y^2 = -\ell^2$ and embedded metric is given by
$ds^2 = -du^2 -dv^2 +dx^2 +dy^2.$ The non-extremal BTZ black holes can then be obtained by the identifications in $\text{AdS}_3$ by the Killing vector
\begin{equation}
    \xi = \frac{r_+}{\ell}J_{ux} - \frac{r_-}{\ell}J_{vy}
\end{equation}
where $J_{ab}$ are the Killing vectors of $\text{AdS}_3$, originating from the $so(2,2)$ algebra. Taking the flat limit and relabelling the coordinates i.e. $\ell \rightarrow \infty\,, v/\ell\to-1\,, u\to T\,, x\to X\,~ \text{and}\,~ y\to Y$, the Killing vector becomes 
\begin{eqnarray}
    \xi_{\small{\text{FSC}}}=\hat{r}_+(X\partial_T+T\partial_X)+r_0\partial_Y.
\end{eqnarray}
The first term generates the boost with velocity $v=\tanh{2\pi\hat{r}_+}$ along the $X$ direction, whereas the second term generates a translation of $2\pi r_0$ along the $Y$ direction. Hence these are known as the shifted-boost orbifolds of $3$D Minkowski space. To establish the connection with FSCs, one makes coordinate transformations as given in table \ref{coordchart} below.
\begin{table}[H]
\centering
\begin{tabular}{|c|c|}
\hline
   Region I ($r>r_0$)  & Region II ($r<r_0$)  \\
\hline\hline
  $\spa{0.4}\displaystyle{T=\sqrt{\frac{r^2-r_0^2}{\hat{r}_+^2}}\cosh{\hat{r}_+\phi}}\spa{-0.7}$  &  $\displaystyle{T=\sqrt{\frac{r_0^2-r^2}{\hat{r}_+^2}}\sinh{\hat{r}_+\phi}}$  \\
\hline
  $\spa{0.4}\displaystyle{X=\sqrt{\frac{r^2-r_0^2}{\hat{r}_+^2}}\sinh{\hat{r}_+\phi}}\spa{-0.7}$  &  $\displaystyle X=\sqrt{\frac{r_0^2-r^2}{\hat{r}_+^2}}\cosh{\hat{r}_+\phi}$  \\
\hline
  $\spa{0.2}Y=r_0\phi-\hat{r}_+t\spa{-0.4}$  &  $Y=r_0\phi-\hat{r}_+t$  \\
\hline
\end{tabular}
 \caption{Coordinate parametrization in two regions}
\label{coordchart}
\end{table}

It can be easily checked under this transformation that $\xi_{\small{FSC}}= \partial_{\phi}$, i.e., the orbifold direction is along $\phi$. Next to proceed with our analysis, in each region we can choose the following coordinates 
\begin{subequations}
\begin{align}
\tau^2&=T^2-X^2=\frac{r^2-r_0^2}{\hat{r}_+^2},\ y=Y,\ x=\hat{r}_+t\quad(r>r_0)\\
\rho^2&=X^2-T^2=\frac{r_0^2-r^2}{\hat{r}_+^2},\ y=Y,\ x=\hat{r}_+t\quad(r<r_0). 
\end{align}
\end{subequations}
We also introduce light-cone coordinates as below
\begin{equation}
X^{\pm}=\frac{T\pm X}{\sqrt{2}}=\begin{cases}
\frac{\tau}{\sqrt{2}}e^{\pm E(x+y)},&\quad(r>r_0)\\
\pm\frac{\rho}{\sqrt{2}}e^{\pm E(x+y)},&\quad(r<r_0)
\end{cases}\quad \text{and}\quad E=\frac{\hat{r}_+}{r_0}.
\end{equation}
Under this choice of the coordinates, the metric takes the following familiar form
\begin{subequations}\label{met}
\begin{align}
    ds^2=dy^2-d\tau^2+E^2\tau^2(dy+dx)^2\,,\quad r>r_0\label{metI}\\
    ds^2=-E^2\rho^2(dy+dx)^2+d\rho^2+dy^2\,,\quad r<r_0.\label{metII}
\end{align}
\end{subequations}
We can interpret the above metrics as the 3D Kaluza-Klein theory with $y$ as the extra compact dimension with identification $y\sim y\pm2\pi r_0$ (since $\xi=r_0\partial_y$). The Kaluza Klein ansatz for 3D is
\begin{equation}
ds_{(3)}^2=ds_{(2)}^2+\Phi^2(dx^3+A_\mu dx^\mu)^2.
\end{equation}
For region I, the metric (\ref{metI}) can be written as
\begin{equation}\label{regI}
    ds^2=-d\tau^2+\frac{(E\tau)^2}{\Phi^2}dx^2+\Phi^2(dy+A_xdx)^2
\end{equation}
where $\Phi^2=1+(E\tau)^2$ and $A_x=1-\Phi^{-2}$. Therefore for $\tau>0$, it represents an expanding universe. Whereas for $(E\tau)^2<<1$, it yields the Milne metric $ds_{(2)}^2=-d\tau^2+(E\tau)^2dx^2$.\\
Similarly, for region II, we can write equation (\ref{metII})  as
\begin{equation}\label{regII}
    ds^2=-\frac{(E\rho)^2}{\Phi^2}dx^2+d\rho^2+\Phi^2(dy+A_xdx)^2\,,
\end{equation}
with $\Phi^2=1-(E\rho)^2$. For $(E\rho)^2<<1$, it yields the Rindler metric $ds_{(2)}^2=-(E\rho)^2dx^2+d\rho^2$. Looking at the metrics (\ref{metI}) and (\ref{metII}) we see that $\xi$ and $\partial_x$ are the two Killing vectors. In the region I, both of these are spacelike. Thus FSCs have no timelike Killing vector in the region I.

\subsection{FSC: Thermodynamics}
Before we head into attempting an understanding of perturbations of the FSC spacetime, let us provide a brief description of the thermodynamics associated with this cosmological solution. 

\medskip

The FSC is endowed with a cosmological horizon that sits at $r=r_0$ (see \eqref{fsc}). This is a Killing horizon and there is an associated surface gravity and hence a temperature that can be attributed to this horizon \cite{Bagchi:2012xr}. This is given by
\begin{align}
    T_{FSC} = \frac{\hat{r}_+^2}{2\pi r_0}\,.
\end{align}
One can also associate an entropy with this horizon. 
\begin{align}
    S_{FSC} = \frac{\text{Area of horizon}}{4\pi G} = \frac{2 \pi r_0}{4\pi G} = \frac{r_0}{2G} = \frac{J}{\sqrt{2GM}}.
\end{align}
The thermodynamic quantities lead to a first law for the FSC \cite{Bagchi:2012xr}. This is given by 
\begin{align}
dM = -T_{FSC} \, dS_{FSC} + \Omega_{FSC} \, dJ
\end{align}
where $\Omega_{FSC}= \frac{\hat{r}_+}{r_0}$ is the angular velocity of FSC. Notice the ``wrong'' sign in front of the $TdS$ term. This is a signature of the fact that the horizon is a remnant of the inner BTZ horizon which also carries a similar unfamiliar sign.

\medskip

The FSC entropy can be reproduced by a counting of states in the dual field theory which is governed by 2D conformal Carrollian symmetry \cite{Bagchi:2012xr}. The entropy of the FSC can also be derived from a limit of the a so-called ``inner-horizon'' Cardy formula on the dual side \cite{Riegler:2014bia, Fareghbal:2014qga}. 

\medskip

It is well known that in AdS spacetimes there exists a phase transition known as the Hawking-Page transition \cite{Hawking:1982dh}. In the context of AdS$_3$, this is a transition between the BTZ black hole and thermal AdS. There exists a flat space analog of this transition where the transition is between hot flat space and the FSC solution \cite{Bagchi:2013lma}. Hence there is a phase transition between flat space and a time evolving cosmological solution at sufficiently high temperatures.

\section{Scalar fields in FSC background}\label{scalar}

In this section, we study the properties of a massive scalar field in the FSC background. In 3D, the scalar wave equation can be exactly solved. We study these solutions and determine conditions that ensure these solutions are well-behaved. We then determine the quasi-normal modes for the scalar field.

The scalar wave equation that we need to solve is the usual
\begin{equation}
    (\Box-m^2)\Psi=0\,.
\end{equation}
where  $m$ is the mass of the scalar field and the covariant Laplacian is given by $\Box=\frac{1}{\sqrt{-g}}\partial_a\left(\sqrt{-g} g^{ab}\partial_b\right).$
Using the metric in equations \eqref{metI} and \eqref{metII}, the wave equation leads to the following second-order differential equations
\begin{subequations}
\begin{align}
    &\left[\tau^2\partial_\tau^2+\tau\partial_\tau-\frac{1}{E^2}\partial_x^2-\tau^2(\partial_y-\partial_x)^2+\tau^2m^2\right]\Psi=0\,,\quad r>r_0,  \label{eqI} \\
    & \left[\rho^2\partial_\rho^2+\rho\partial_\rho-\frac{1}{E^2}\partial_x^2+\rho^2(\partial_y-\partial_x)^2-m^2\rho^2\right]\Psi=0\,,\quad r<r_0.\label{eqII}
\end{align}
\end{subequations}
We now search for a wave function which solves the above equation and is invariant under the action of the discrete subgroup generated by $\xi$ i.e. $\Psi(e^{2\pi n\xi}X)=\Psi(X)\,, n\in\mathbb{Z}$. As is evident from the FSC metric (\ref{metI}), $\xi$ and $\partial_x$ are two commutating Killing vectors, we therefore choose their simultaneous eigenfunctions $\psi_{n,p}$ as the basis to mode expand $\Psi$ as follows
\begin{align}
\xi\psi_{n,p}&=in\psi_{n,p}\,, n\in\mathbb{Z} ~~~~~\mbox{and}~~~~~
\partial_x\psi_{n,p}=ip\psi_{n,p}\,, p\in\mathbb{C}\,.
\end{align}
This inspires us to consider the following ansatz for our solution
\begin{equation}
    \psi_{n,p}^{I}=f^{I}(\tau)e^{ipx}e^{i\frac{n}{r_0}y}
~~~~~~\mbox{and}~~~~~
    \psi_{n,p}^{II}=f^{II}(\rho)e^{ipx}e^{i\frac{n}{r_0}y},
\end{equation}
which after substitution in equations \eqref{eqI},\eqref{eqII} leads to  the following differential equations
\begin{subequations}
\begin{align}
&\left[\tau^2\frac{d^2}{d\tau^2}+\tau\frac{d}{d\tau}+(\omega^2\tau^2-\nu^2)\right]f^I(\tau)=0, \label{eqI2}\\
&\left[\rho^2\frac{d^2}{d\rho^2}+\rho\frac{d}{d\rho}+((i\omega)^2\rho^2-\nu^2)\right]f^{II}(\rho)=0 \label{eqII2}
\end{align}
\end{subequations}
where $\omega^2=(p-n/r_0)^2+m^2$ and $\nu=i\frac{p}{E}$. The solutions to these are given in terms of the Bessel functions and are written as
\begin{equation}
    \psi_{n,p}^{I,\pm}=J_{\pm\nu}(\omega|\tau|)e^{ipx}e^{i\frac{n}{r_0}y} ~~~~\mbox{and}~~~~~
    \psi_{n,p}^{II,\pm}=J_{\pm\nu}(i\omega|\rho|)e^{ipx}e^{i\frac{n}{r_0}y}\,.
\end{equation}
For non-integer $\nu$, $\psi^{\pm}$ are two independent solutions. Consider the equation \eqref{eqI2}. We can also write this differential equation in Schr\"odinger-like form which gets rid of the first-order term from the equation as
\begin{align}
&\left[\frac{d^2}{\ d\tau^{2}_*}+(\omega^2 e^{2\tau_*}-\nu^2)\right]f^I(\tau_*)=0
\end{align}
where $\tau\dfrac{d}{d\tau}=\dfrac{d}{d\tau_*}\ \Rightarrow\ \tau_*=\ln{\tau}$. The effective potential experienced by this scalar particle thus is
\begin{equation}
V_{\text{eff}}(\tau_*)=-\omega^2 e^{2\tau_*}.
\end{equation}

\subsection{Wavefunctions}\label{fscwf}
For the purpose of this paper, we analyze the solutions in the region $\text{I}_{out}$, where $\tau>0$. In this region, our solution is 
\begin{equation} \label{oursol}
    \psi_{n,p}^{\pm}=J_{\pm\frac{ip}{E}}(\omega\tau)e^{ipx}e^{i\frac{n}{r_0}y}.
\end{equation}
We demand the solutions to satisfy the following conditions:
\begin{enumerate}
    \item The solutions are bounded at spatial infinities $|x|\to\infty$.
   \item The solutions are regular at the cosmological horizon $\tau=0$.
\end{enumerate}
These conditions are imposed in the manner described below.
Generally, one considers the Bessel function of the second kind to be the other independent solution of the differential equation \eqref{eqI2}. However, the Bessel function of the second kind diverges for $\tau\to0$ and thus is discarded.
Working thus with the solution in \eqref{oursol}, if we consider $p$ as momentum along the $x$ direction and initially assume it to be real, it is evident that the frequency $\omega$ is real-valued and we would have two solutions, as given below
\begin{eqnarray}
  \hspace{-0.5cm}  \psi_{n,p}^{I,\pm}=J_{\pm i\frac{p}{E}}(\omega\tau)e^{ipx}e^{i\frac{n}{r_0}y}
    =e^{ip\left(\pm\frac{\ln\tau}{E}+x\right)}e^{i\frac{n}{r_0}y}\left(\frac{\omega}{2}\right)^{\frac{ip}{E}}\sum_{s=0}^\infty\frac{(-1)^s}{s!\ \Gamma(s+1\pm ip/E)}\left(\frac{\omega\tau}{2}\right)^{2s}
\end{eqnarray}
where, looking at the phase or the near horizon behavior, we infer that $\psi^+$ is the \textit{in-going} while $\psi^-$ is the \textit{out-going} solution. 
In our search for quasi-normal modes, we will promote $p$ to be a complex number, $p=p_r+ip_i$ which gives 
\begin{equation}
    \psi^{\pm}_{n,p}=e^{-p_ix}\left(\frac{\omega\tau}{2}\right)^{\mp\frac{p_i}{E}}\left[\sum_{s=0}^{\infty}\frac{(-1)^s}{s!\Gamma\left(s+1\mp\frac{p_i}{E}\pm i\frac{p_r}{E}\right)}\left(\frac{\omega\tau}{2}\right)^{2s\pm i\frac{p_r}{E}}\right]e^{ip_rx}e^{i\frac{n}{r_0}y}.
\end{equation}
We will see that this generates complex valued $\omega$'s. We note the following points regarding the behavior of the scalar field solution:
\begin{itemize}
    \item For $x>0$, we need to restrict ourselves to $p_i>0$. Under these conditions $\psi^-_{n,p}$ is well-behaved however $\psi^+_{n,p}$ diverges as $\tau\to0$ for all $x$, unless the gamma function also diverges \textit{i.e.} when $1-\frac{p_i}{E}+i\frac{p_r}{E}=1-N,\ N\in\mathbb{N}$. This yields $p_r=0$ and $p_i=EN$. 
    The surviving function and $\psi^-_{n,p}$ are both well-behaved at $\tau\to0$ and are written as
\begin{eqnarray}
    \psi_{n,N}^{+}&=&e^{-ENx}\left[\left(\frac{\omega\tau}{2}\right)^{N}\sum_{s=0}^{\infty}\frac{(-1)^{s+N}}{(s+N)!\Gamma\left(s+1\right)}\left(\frac{\omega\tau}{2}\right)^{2s}\right]e^{i\frac{n}{r_0}y}=(-1)^Ne^{-ENx}J_{N}(\omega\tau)e^{i\frac{n}{r_0}y}\nonumber\\
    \psi_{n,N}^-&=&e^{-ENx}J_{N}(\omega\tau)e^{i\frac{n}{r_0}y}.
\end{eqnarray}

    \item Similarly, for $x<0$, we need $p_i<0$. Now, $\psi^-_{n,p}$ diverges while  $\psi^+_{n,p}$ is well behaved. Following the same arguments as above, we obtain $p_r=0$ and $p_i=-EN$. The solutions now are given by
    \begin{eqnarray}
    \psi_{n,N}^{-}&=&e^{ENx}\left[\left(\frac{\omega\tau}{2}\right)^{N}\sum_{s=0}^{\infty}\frac{(-1)^{s+N}}{(s+N)!\Gamma\left(s+1\right)}\left(\frac{\omega\tau}{2}\right)^{2s}\right]e^{i\frac{n}{r_0}y}=(-1)^Ne^{ENx}J_{N}(\omega\tau)e^{i\frac{n}{r_0}y}\nonumber\\
    \psi_{n,N}^+&=&e^{ENx}J_{N}(\omega\tau)e^{i\frac{n}{r_0}y}.
\end{eqnarray}
\end{itemize}
This also dictates our choice of boundary condition at the cosmological horizon which we impose in the next subsection. From the above analysis it is clear that we cannot impose either the ingoing or the outgoing boundary conditions, as the fate of the ingoing and outgoing waves are intertwined with each other. Imposing either one to die off also trivially eliminates the other as is detailed in appendix \ref{inout}.

However, to have well-behaved solutions, we need to pay a price. The solutions are no longer independent. 
Thus, effectively we have two solutions $(N\in\mathbb{N})$
\begin{equation}\label{sol}
\begin{aligned}
\psi^{x>0}_{n,N}&=e^{-ENx}J_N(\omega\tau)e^{i\frac{n}{r_0}y} ~~~,~~~
\psi^{x<0}_{n,N}&=e^{ENx}J_N(\omega\tau)e^{i\frac{n}{r_0}y}.
\end{aligned}
\end{equation}
Further, the asymptotic behavior of the Bessel function for large $\tau$ is given as
\begin{equation}\label{Jlarge}
    J_{N}(\omega\tau)\sim\sqrt{\frac{2}{\omega\pi\tau}}\cos{\left(\omega\tau-\frac{2N+1}{4}\pi\right)}.
\end{equation}
Hence, the solutions in \eqref{sol} are oscillatory in $\tau$ but decaying along the $x$ direction. We also note that it is also not differentiable at $x=0$ because they show a cusp at this $x$ value as depicted in figure \ref{f(x)solution}.
\begin{figure}[H]
\centering
\begin{subfigure}{.5\textwidth}
  \centering
  \includegraphics[width=.8\linewidth]{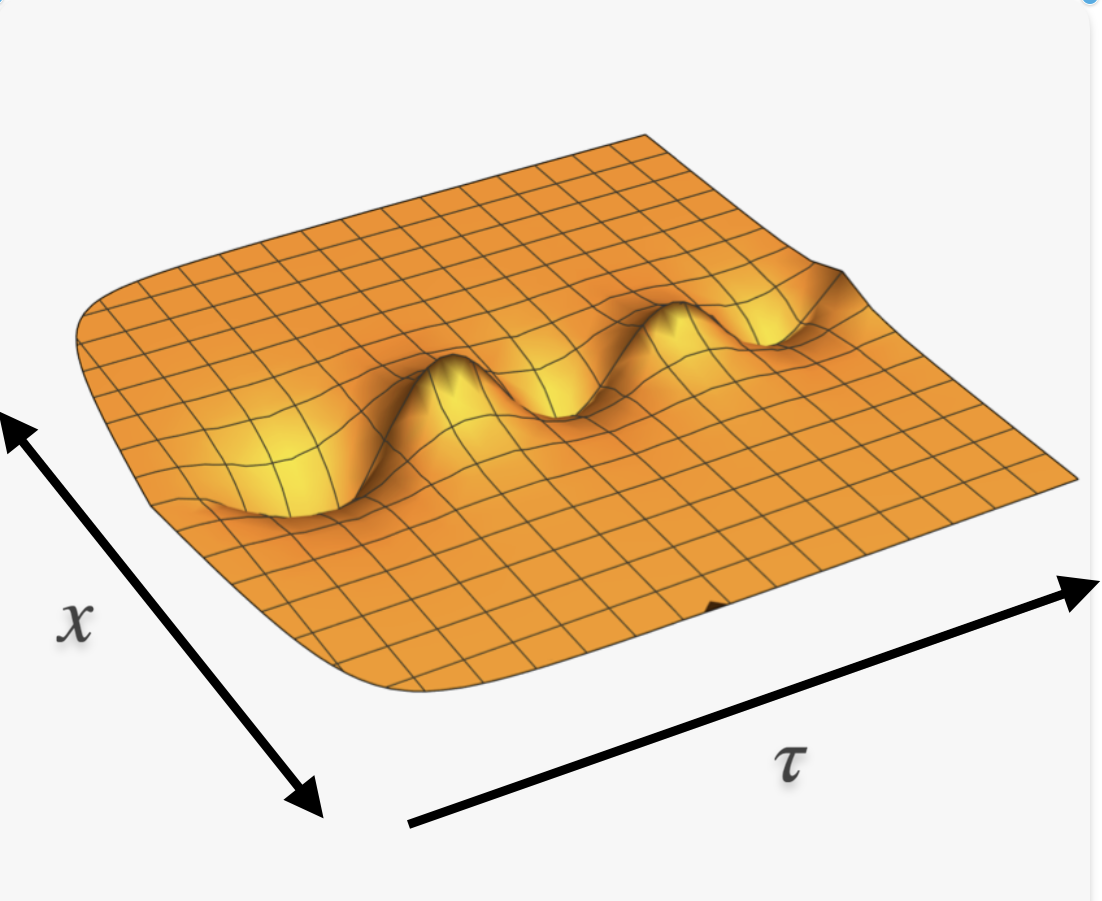}
  \label{fig:sub1}
\end{subfigure}%
\begin{subfigure}{.5\textwidth}
  \centering
  \includegraphics[width=.9\linewidth]{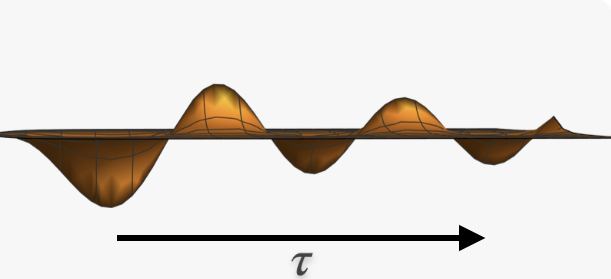}
  \label{fig:sub2}
\end{subfigure}
\caption{Representative decaying solution as defined in \eqref{gensol}.}
\label{f(x)solution}
\end{figure}

\noindent Thus, we can write a general solution as follows:
\begin{equation} \label{gensol}
\Psi=\underbrace{J_{\pm i\frac{p}{E}}(\omega_p\tau)e^{ipx}e^{i\frac{n}{r_0}y}}_{\text{oscillatory along }x}+\underbrace{e^{-EN|x|}J_N(\omega_{{}_N}\tau)e^{i\frac{n}{r_0}y}}_{\text{decaying along }x}
\end{equation}

Here, $\omega_p$ is real-valued and $\omega_N$ is complex-valued frequency. This solution is now well behaved and we may proceed with our analysis. The oscillatory solutions were already known \cite{Cornalba_2004}. However, the decaying solutions, which we obtain by complexifying $p$, are new.

\begin{figure}[H] 
\begin{center} 
    \includegraphics[width=0.5\linewidth]{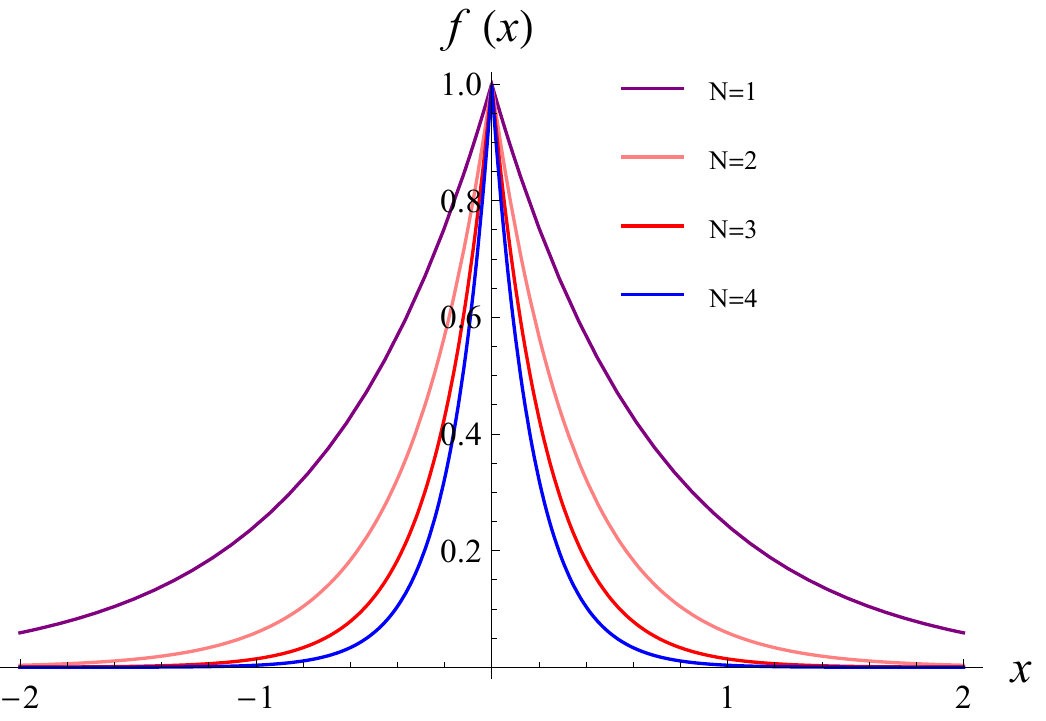} 
    \caption{Representative decaying solution as defined in \eqref{gensol} with $M=2, J=1$.} 
    \label{f(x)solution} 
  \end{center} 
\end{figure}

\subsection{Reflective boundary condition}\label{rbc}
Dictated by the considerations in the previous subsection, we now impose reflective boundary conditions or a ``hard wall'' at the cosmological horizon. The general solution for $\tau>0$ can be expressed as
\begin{align}
    \psi_{n,p}^{I,\pm}&=J_{\pm i\frac{p}{E}}(\omega\tau)e^{ipx}e^{i\frac{n}{r_0}y}
    =e^{ip\left(x\pm\frac{1}{E}\ln(\frac{\omega\tau}{2})\right)}e^{i\frac{n}{r_0}y}\sum_{s=0}^\infty\frac{(-1)^s}{s!\ \Gamma(s+1\pm ip/E)}\left(\frac{\omega\tau}{2}\right)^{2s}\nonumber\\
    &=\frac{e^{ip\left(x\pm\frac{\tau_*}{E}\right)}e^{\pm i\frac{\ln(\omega/2)}{E}}e^{i\frac{n}{r_0}y}}{\Gamma(1\pm ip/E)}\ F\left(1\pm i\frac{p}{E};-\frac{\omega^2\tau^2}{4}\right)
\end{align}
where $p\in\mathbb{C}$ and,
\begin{equation}
F\left(1\pm i\frac{p}{E};-\frac{\omega^2\tau^2}{4}\right)=\sum_{s=0}^\infty\frac{\Gamma(1\pm i\frac{p}{E})}{s!\ \Gamma(s+1\pm i\frac{p}{E})}\left(-\frac{\omega^2\tau^2}{4}\right)^{s}.
\end{equation}
As discussed earlier, $e^{ip(x\pm\frac{\tau_*}{E})}$ indicate ingoing and outgoing waves at the horizon. Thus $\psi^+$ is the \textit{ingoing} and $\psi^-$ is the \textit{outgoing} wave at the horizon. Now, we calculate the flux for this solutions.
\begin{align*}\label{flux}
\mathcal{F}^{\pm}_{n,p}&=-\sqrt{-g}\ \text{Im}(\psi^{\pm\,*}_{n,p}\partial_\tau\psi^{\pm}_{n,p})\\
&=-E\tau\ \text{Im}\left[\left(J_{\pm i\frac{p}{E}}(\omega\tau)\right)^*\partial_\tau J_{\pm i\frac{p}{E}}(\omega\tau)e^{-ip^*x}e^{ipx}\right]\\
&=-E\tau\,e^{-2p_i x}\ \text{Im}\left[\left(J_{\pm i\frac{p}{E}}(\omega\tau)\right)^*\left(-\omega J_{1\pm i\frac{p}{E}}(\omega\tau)\pm\frac{ip}{E\tau}J_{\pm i\frac{p}{E}}(\omega\tau)\right)\right]\\
&=E\tau\,e^{-2p_i x}\ \text{Im}\left[\omega\left(J_{\pm i\frac{p}{E}}(\omega\tau)\right)^* J_{1\pm i\frac{p}{E}}(\omega\tau)\right]\mp p_r\,e^{-2p_i x}\ \left|J_{\pm i\frac{p}{E}}(\omega\tau)\right|^2
\end{align*}
The reflective boundary condition demands
\begin{equation}
\mathcal{F}^+_{n,p}=-\mathcal{F}^-_{n,p}\ \Rightarrow\ \mathcal{F}^+_{n,p}+\mathcal{F}^-_{n,p}=0.
\end{equation}
The above translates to the following condition
\beqa
 \mathcal{F}^+_{n,p}+\mathcal{F}^-_{n,p}&=&E\tau\,e^{-2p_i x}\ \text{Im}\left[\omega\left(J_{i\frac{p}{E}}(\omega\tau)\right)^* J_{1+i\frac{p}{E}}(\omega\tau)+\omega\left(J_{-i\frac{p}{E}}(\omega\tau)\right)^* J_{1-i\frac{p}{E}}(\omega\tau)\right]\nonumber\\
&&-p_r\,e^{-2p_i x}\ \left|J_{i\frac{p}{E}}(\omega\tau)\right|^2+p_r\,e^{-2p_i x}\ \left|J_{-i\frac{p}{E}}(\omega\tau)\right|^2=0.
\eeqa
First we put $p_r=0$ which means $i\frac{p}{E}=-\frac{p_i}{E}$, which gives us
\beqa
 \mathcal{F}^+_{n,p}+\mathcal{F}^-_{n,p}&=&E\tau e^{-2p_i x}\text{Im}\left[\omega\frac{\left(\frac{\omega^*\tau}{2}\right)^{-\frac{p_i}{E}}}{\Gamma(1-\frac{p_i}{E})}\frac{\left(\frac{\omega\tau}{2}\right)^{1-\frac{p_i}{E}}}{\Gamma(2-\frac{p_i}{E})}+\omega\frac{\left(\frac{\omega^*\tau}{2}\right)^{\frac{p_i}{E}}}{\Gamma(1+\frac{p_i}{E})}\frac{\left(\frac{\omega\tau}{2}\right)^{1+\frac{p_i}{E}}}{\Gamma(2+\frac{p_i}{E})}\right] \nonumber\\
&=&\ e^{-2p_i x}\text{Im}\left[\frac{\frac{\omega^2\tau^2}{2}\left(\frac{|\omega|\tau}{2}\right)^{-\frac{2 p_i}{E}}}{\Gamma(1-\frac{p_i}{E})\Gamma(2-\frac{p_i}{E})}+\frac{\frac{\omega^2\tau^2}{2}\left(\frac{|\omega|\tau}{2}\right)^{\frac{2 p_i}{E}}}{\Gamma(1+\frac{p_i}{E})\Gamma(2+\frac{p_i}{E})}\right]=0. 
\eeqa
Since we also have
$\omega^2=(\omega_r+i\omega_i)^2=\omega^2_r-\omega^2_i+i2\omega_r\omega_i$, our demand in the previous equation reduces to the following condition
\begin{equation}
 e^{-2p_i x}\omega_r\omega_i\left[\left(\frac{|\omega|}{2}\right)^{-\frac{2 p_i}{E}}\frac{\tau^{2(1-\frac{p_i}{E})}}{\Gamma(1-\frac{p_i}{E})\Gamma(2-\frac{p_i}{E})}+\left(\frac{|\omega|}{2}\right)^{\frac{2 p_i}{E}}\frac{\tau^{2(1+\frac{p_i}{E})}}{\Gamma(1+\frac{p_i}{E})\Gamma(2+\frac{p_i}{E})}\right]=0.
\end{equation}
We now note the following. If $1-\frac{p_i}{E}<0\ \Rightarrow \frac{p_i}{E}>1$, then $\tau^{2(1-\frac{p_i}{E})}$ diverges unless $\Gamma(1-\frac{p_i}{E})=\infty$. This happens when $1-\frac{p_i}{E}=-N,\ N\in\mathbb{N}$ \textit{i.e.} $p_i=E(1+N)$. When $\frac{p_i}{E}=1$ the vanishing of flux is still satisfied due to the $\Gamma(1-\frac{p_i}{E})$ factor in denominator of the first term as $\Gamma(0)=\infty$. So, for $p_i>0$ we have $p_i=EN,\ N\in\mathbb{N}$. Similarly, for $\frac{p_i}{E}<-1$, $\tau^{2(1+\frac{p_i}{E})}$ diverges unless $\Gamma(1+\frac{p_i}{E})=\infty$. As also indicated in the previous subsection, we thus can now conclude that our boundary conditions require $p_i=-EN,\ N\in\mathbb{N}$. Finally, for $p_i=0$, the boundary condition is satisfied by the $\tau^2$ factor. Thus, we get $p_i=EN,\ N\in\mathbb{Z}$.

\subsection{Quasi-normal modes}\label{qnmcalc}
Having found wavefunctions respecting the reflective boundary condition, we proceed to compute the quasi-normal modes for scalar perturbation in the FSC background. As only $\omega_N$'s can be considered quasi-normal, we drop the subscript. From our previous analysis, we found the following expression for frequency in terms of momentum and mass of the scalar field
\begin{equation}
\omega^2=\left(p_r-\frac{n}{r_0}+ip_i\right)^2+m^2=\left[-\frac{n}{r_0}+i\frac{\hat{r}_+}{r_0}N\right]^2+m^2.
\end{equation}
This gives us the following relations for the real and imaginary components of the frequency
\begin{eqnarray}
\ &\omega^2_r-\omega^2_i+i2\omega_r\omega_i=\frac{n^2}{r_0^2}-\frac{\hat{r}_+^2}{r_0^2}N^2+m^2-i\frac{2\hat{r}_+}{r_0^2}nN\nonumber\\
\Rightarrow\ &\omega^2_r-\omega^2_i=\frac{n^2}{r_0^2}-\frac{\hat{r}_+^2}{r_0^2}N^2+m^2;\quad\omega_r\omega_i=-\frac{\hat{r}_+}{r_0^2}nN.
\end{eqnarray}
We denote $P_n=-\frac{n}{r_0}$ and $Q_N=\frac{\hat{r}_+}{r_0}N$. Thus, $\omega^2_r-\omega^2_i=P_n^2-Q_N^2+m^2$, $\omega_r\omega_i=P_nQ_N$. After some algebraic simplification, we find that $\omega_i$ satisfies the following equation
\begin{equation}
    \omega_i^4+(P_n^2-Q_N^2+m^2)\omega_i^2-P_n^2Q_N^2=0,
\end{equation}
which after solving gives us the following 
\begin{align}
    &\omega_{i,r}^2=\frac{1}{2}\left[\pm\sqrt{(P_n^2-Q_N^2+m^2)^2+4P_n^2Q_N^2}\mp (P_n^2-Q_N^2+m^2)\right]
\end{align}    
We observe that $N=0$ should correspond to the normal modes. This can be seen as follows. As we are considering a relativistic theory, the normal modes must be $\omega_r^2=P_n^2+m^2$. If we choose the sign convention to be `+', then this is indeed the case. Thus, we get $\omega^2_i=0$ and $\omega_r^2=P_n^2+m^2$. Let us now take the momentum to be zero \textit{i.e.} $n=0$. Under this convention $\omega^2_r=m^2-Q^2_N$ and $\omega^2_i=0$. The $\omega^2_i$ survives only when both $n$ and $N$ are non-zero. Thus we get the following expressions for the imaginary and real components of the  quasi-normal frequencies
\begin{align}
\boxed{\omega_{i,r}^2=\sqrt{A_{n, N}^2+ B_{n,N}^2}\mp A_{n,N}, \, \text{where}\,\, A_{n,N}= \frac{n^2-N^2\hat{r}_+^2 +m^2r_0^2}{2 r_0^2}, \, B_{n,N}=\frac{\hat{r}_+nN}{r_0}.}
\end{align}
The above are thus the expressions for QNM associated with scalar perturbations of an FSC solution. This is one of the main results of our paper. 

\subsection{Wave Function and Effective Potential from Limiting analysis}\label{limit}
We have stressed that the FSC can also be obtained by taking the flat-space limit ($\ell\to\infty$) from the BTZ black hole. In this subsection, we will perform a flat space limit over the computations for scalar fields in the BTZ background and attempt to connect to our analysis which till now was purely intrinsic, i.e. was in 3D AFS without any reference to 3D AdS spacetimes and analyses there.  

We study this limit here for scalar fields in the intermediate region of the BTZ black hole. This also serves as a consistency check for our computations. To do so, one solves for the scalar wave equation and gets certain constraints on the parameters that characterize our solutions as described in Appendix \ref{APPA}.
We then find out what happens to the solutions in the flat space limit. We take the $\ell\to\infty$  limit on the coordinates and the scaling dimension of the field as follows,
\begin{align}
&\frac{\Delta}{\ell}\rightarrow m~,~\ell^2z \rightarrow \frac{r^2-r_0^2}{\hat{r}_+^2}=\tau^2~,~ x^+\rightarrow \hat{r}_+t-r_0\varphi=-y~,~\frac{x^-}{\ell}\rightarrow \hat{r}_+\varphi=E(y+x)\,.
\end{align}
Then we consider how the flat space limit acts on the following combination $k_+x^++k_-x^-$
\begin{eqnarray}
\lim_{\ell\to\infty}(k_+x^++k_-x^-)=k_-\ell Ex+(k_-E\ell-k_+)y\,.
\end{eqnarray}
As the phase in the exponential must match with FSC solutions, we demand
\begin{eqnarray}
k_-\ell Ex+(k_-\ell E-k_+)y=-px-\frac{n}{r_0}y.
\end{eqnarray}
Now, comparing both sides, we get
\begin{eqnarray}\label{limaid}
k_-\ell\to -\frac{p}{E},\quad k_+\to\left(\frac{n}{r_0}-p\right)
\end{eqnarray}
and
\begin{eqnarray}
a\to\frac{\ell}{2}\left[m-i\left(\frac{n}{r_0}-p\right)\right],\ b\to-\frac{\ell}{2}\left[m+i\left(\frac{n}{r_0}-p\right)\right],\ c\to1+i\frac{p}{E}.
\end{eqnarray}
The first point to note is $k_-\ell=iEN,\ N\in\mathbb{Z}$ implies $p=iE(-N)$. Thus, we successfully retrieve our result in the limiting analysis. We can also map the hypergeometric function, which appears in the BTZ solutions, to the Bessel function in the flat space limit, 
\begin{eqnarray}
\lim_{\ell\to\infty}\frac{{}_2F_1(a,b;c;z)}{\Gamma(c)}&=&\sum_{s=0}^{\infty}\frac{z^s}{s!\Gamma(1+ip/E)}\left[\prod_{j=0}^{s-1}\frac{-\frac{\ell^2}{4}\left[m-i\left(\frac{n}{r_0}-p\right)\right]\left[m+i\left(\frac{n}{r_0}-p\right)\right]}{\left(j+1+i\frac{p}{E}\right)}\right]\nonumber\\
&&=\sum_{s=0}^{\infty}\frac{\left[m^2+\left(\frac{n}{r_0}-p\right)^2\right]^s}{s!\Gamma\left(s+1+i\frac{p}{E}\right)}\left(-\frac{\ell^2z}{4}\right)^s\nonumber\\
&&=\sum_{s=0}^{\infty}\frac{(-1)^s}{s!\Gamma\left(s+1+i\frac{p}{E}\right)}\left(\frac{\omega\tau}{2}\right)^{2s}=\left(\frac{\omega\tau}{2}\right)^{-ip/E}J_{\frac{ip}{E}}(\omega\tau).
\end{eqnarray}
Similarly, we have
\begin{equation}
1+a-c\to\frac{\ell}{2}\left[m-i\left(\frac{n}{r_0}-p\right)\right],\ 1+b-c\to-\frac{\ell}{2}\left[m+i\left(\frac{n}{r_0}-p\right)\right],\ 2-c\to1-i\frac{p}{E}.
\end{equation}
Thus,
\begin{equation}
\lim_{\ell\to\infty}\frac{{}_2F_1(1-a+c,1-b+c;2-c;z)}{\Gamma(2-c)}=\left(\frac{\omega\tau}{2}\right)^{ip/E}J_{-\frac{ip}{E}}(\omega\tau).
\end{equation}
Thus, the wave functions reduce to our FSC wave functions up to a constant factor 
\begin{equation}
\lim_{\ell\to\infty}\ell^{N}\psi^{\pm}_{N}=\left(\frac{2}{\omega}\right)^{N} e^{-EN|x|}e^{i\frac{n}{r_0}y}J_{N}(\omega\tau).
\end{equation}
Next we look for the limiting analysis for the effective potential. The differential equation satisfied by a scalar in the intermediate region ($0\leq z \leq 1$) of BTZ is
\begin{equation}
    z(1-z)\frac{d^2R}{dz^2}+(1-2z)\frac{dR}{dz}+\left[\frac{\ell^2k_+^2}{4(1-z)}+\frac{\ell^2k_-^2}{4z}+\frac{m^2\ell^2}{4}\right]R(z)=0.
\end{equation}
We can again bring this equation to the Schroedinger-like form by making a change of variables as $ z_{\star}=\ln\f{z}{1-z}$.
This gives
\beqa
\f{d^2 R}{d z^2_{\star}}+\f{\ell^2}{4}\left[k_+^2+\frac{(k_+^2 - k_-^2)e^{z_\star}}{(e^{z_{\star}}+1)}+\frac{m^2 e^{z_{\star}}}{(e^{z_{\star}}+1)^2}\right]R(z)=0.
\eeqa
Again, the identification in equation \ref{limaid} and $
e^{z_{\star}}=\f{\tau^2}{\ell^2} \rightarrow 0
$
gives us
\begin{align}
    \left[\frac{d^2}{\ d\tau^{2}_*}+\left\lbrace \f{p^2}{E^2}
+\left( \left( \f{n}{r_0}-p\right)^2+m^2\right)\tau^2\right\rbrace\right]f^I(\tau_*)= \left[\frac{d^2}{\ d\tau^{2}_*}+\left(\omega^2 e^{2\tau_*}+\f{p^2}{E^2}\right)\right]f^I(\tau_*)=0.
\end{align}
Thus, we also recover our effective potential and differential equation in the flat space limit.

\section{One-loop partition function for scalar field in FSC}\label{dhs}

We now exploit our quasi-normal frequencies to build one-loop determinants of the scalar field in FSC background following the DHS prescription \cite{Denef:2009kn}. Following initial success for static space-times \cite{Denef:2009yy,Martin:2019flv}, this prescription has also been generalized to spacetimes that are not necessarily static but are stationary \cite{Castro:2017mfj}. Before we delve into the details of our calculation, let's quickly give an overview of this prescription. 

The DHS prescription utilizes the Weierstrass factorization theorem, according to which a meromorphic function, in our case the one-loop partition function $Z^{(1)}(\Delta)$, analytically continued to the complex $\Delta$ plane, may be written as a product of its poles and zeros and upto a function $e^{\text{Pol}(\Delta)}$.
For the case of AdS one relies on the meromorphicity of the determinant in the mass parameter $\Delta(m^2)$, which is later identified as the conformal dimension of the operator in the dual CFT and is given as 
$\Delta(\Delta-d)=m^2$. 
Specializing to the case of a complex scalar field one may thus write
 \beqa
Z^{(1)}\left(\Delta\right)=\int \mathcal{D}\phi\, \exp \left(\int  d^{d+1} x \sqrt{g} \phi^{*} \left(-\Box+m^2\right)\phi\right)=\text{det}\left(-\Box+\Delta(\Delta-d)\right)^{-1}
\eeqa
where $\Box$ is the scalar Laplacian. 
In this case there are no zeros and the poles occur whenever $\Delta$ is tuned to give a zero mode of the scalar field. These zero modes are such solutions of the Euclidean equation of motion that are smooth and respect the periodicity condition in the Euclidean time direction.

As one Wick rotates to the Lorentzian point of view, the solutions to the Klein-Gordon equation that satisfy ingoing or outgoing boundary conditions at the black hole horizon and normalizable asymptotic boundary conditions are given by these quasi-normal and anti-quasi-normal modes respectively. This prescription can also be generalized to non-static space-times by modifying the regularity condition in the Euclidean version. Since, the thermal frequencies get modified to incorporate both the temperature and angular velocity, the regularity condition imposes restrictions on quantum numbers conjugate to both these. The review of the partition function computation for the BTZ black hole has been reviewed in appendix \ref{APPB}.

\medskip

Now let's proceed to apply this method for scalar fields in FSC background. To find the thermal frequencies we demand the regularity of this solution near the horizon. Near the horizon ($\tau\rightarrow 0$) our solutions behave as
\beqa
\psi \sim \tau^{\nu} \ e^{- \nu E x} \ e^{i\f{n}{r_0}y}.
\eeqa
Regularity requires that $\nu=a$ where $a \in \mathbb{Z}$ .
Also, requiring $\nu =a $ fixes the (anti)quasi-normal frequency $\omega$ to have
the specific value $\omega_a$.

Periodicities in $t$ and $\phi$ coordinates demand both the frequency $\omega$ and wave number $k$ to take integer values. This can be seen as follows. From the FSC metric in equation (\ref{fsc}), we perform the Wick rotations $t=i t_E$ and $\hat{r}_{+}=-i\tilde{r}_{+}$ which yield the Euclidean FSC metric
\beqa
ds_E^2=\f{r^2 dr^2}{\tilde{r}_+^2(r^2-r_0^2)}+\f{\tilde{r}_+^2(r^2-r_0^2)}{r^2}dt_E^2+r^2\left(d\phi-\f{r_0\tilde{r}_+}{r^2}dt_E\right)^2.
\eeqa
Next moving to the near horizon region $r^2=r_0^2+\epsilon \sigma^2$, we get
\beqa
ds_E^2=r_0^2\left(d\phi-\f{\tilde{r}_+}{r_0}dt_E\right)^2+\f{\epsilon}{\tilde{r}_+^2}\left(d\sigma^2+\tilde{r}_+^2\sigma^2d\phi^2\right)^2+\mathcal{O}(\epsilon^2).
\eeqa
We next periodically identify the Euclidean time direction and ensure that the metric is regular by avoiding any conical singularity. This require the transverse direction $\phi-\f{\tilde{r}_+}{r_0}t_E$ to be kept fixed while identifying
\beqa
t_E\sim t_E+\f{2\pi r_0}{\tilde{r}_+^2} ~~,~~\phi \sim \phi+\f{2\pi}{\tilde{r}_+}.
\eeqa
Applying the above we get
\beqa
e^{\f{2\pi}{\tilde{r}_+^2}(-\tilde{r}_+k+r_0\omega)}=1.
\eeqa
The thermal frequency is thus defined as \footnote{
The two coordinate systems are connected as $
e^{i\omega t} e^{-i k \phi}=e^{i\f{\tilde{r}_+a}{r_0}x}e^{i\f{n}{r_0}y}$ which again gives $\omega=i\tilde{r}_+\left(\f{\tilde{r}_{+} a}{r_0}+\f{k}{r_0}\right)$ and $k=-n$. We drop the toleman factor $\hat{r}_+$ for our analysis.

}
\beqa
\omega_a=\f{\tilde{r}_{+} a}{r_0}-\f{n}{r_0}
\eeqa
where $a\in \mathbb{Z}$ is the thermal quantum number and $n$ is the momentum quantum number. \\
\noindent
Using the DHS prescription for non static spacetimes, the scalar one-loop determinant is \cite{Castro:2017mfj}
\beqa
\left(\f{Z^{(1)}(m)}{e^{\text{Pol}(m)}}\right)^2&=&\prod_{a ,*} \left(\omega_a-\omega_{*}\right)^{-1}.
\eeqa
In the above equation, $\star$ represents some set of additional quantum numbers apart from the thermal and momentum quantum number. The boundary conditions and the nature of the quasi-normal modes \cite{Denef:2009kn} are satisfied only at a particular set of discrete frequencies with $\omega^{}_{\star}=\omega^{}_{a}$. 
Here, $\text{Pol}(m)$ is a polynomial in the mass parameter which contains the UV counterterms. 
We also note that in flat space $\Delta = m$. Thus our analyticity arguments are in $m$. Since we want the determinant for a real scalar, hence the square on the left hand side of the above equation. \\

Our quasi-normal frequencies are $\omega_{*}=\omega_r +i\omega_i$ and can be written as

\beqa
\omega_{r,i}^2=\f{m^2}{2}\left[\sqrt{1+\left(\f{P_n^2+Q_N^2}{m^2}\right)^2+\left(\f{2(P_n^2-Q_N^2)}{m^2}\right)}\pm 1\pm \f{P_n^2-Q_N^2}{m^2}\right].
\eeqa
Thus we may write
\beqa
\left(\f{Z^{(1)}(m)}{e^{\text{Pol}(m)}}\right)^{-2}&=&\prod_{a > 0,n,N} \left( \f{i \hat{r}_{+}a}{r_0}-\f{n}{r_0}-\omega_r
-i\omega_i\right)  \left( \f{i \hat{r}_{+}a}{r_0}-\f{n}{r_0}-\omega_r
+i\omega_i \right)\nonumber\\
&\times& \prod_{a > 0,n,N} \left( -\f{i \hat{r}_{+}a}{r_0}-\f{n}{r_0}-\omega_r
-i\omega_i \right) \left( -\f{i \hat{r}_{+}a}{r_0}-\f{n}{r_0}-\omega_r
+i\omega_i \right)\nonumber\\
&\times& \prod_{n,N} \left(-\f{n}{r_0}-\omega_r
-i\omega_i \right)\left(-\f{n}{r_0}-\omega_r
+i\omega_i\right).
\eeqa
The first line corresponds to the ingoing modes with $a > 0$, the second line are the outgoing modes and thermal frequencies
with $a < 0$, and the last line corresponds to the zero modes with $a= 0$. This can be further simplified to give
\beqa
\left(\f{Z^{(1)}(m)}{e^{\text{Pol}(m)}}\right)^{-2}&=&\prod_{a > 0, n,N}\left[\left(\f{n}{r_0}+\omega_r \right)^2+\left(\f{\hat{r}_+ a}{r_0}-\omega_i\right)^2\right]\prod_{a > 0, n,N}\left[\left(\f{n}{r_0}+\omega_r \right)^2+\left(\f{\hat{r}_+ a}{r_0}+\omega_i\right)^2\right] \nonumber\\
&\times& \prod_{n,N} \left[\left(\f{n}{r_0}+\omega_r \right)^2+\omega_i^2\right].
\eeqa
We may now extract out the divergent terms and regulate the product over $a$ to give
\beqa \label{genZ}
\left(\f{Z^{(1)}(m)}{e^{\text{Pol}(m)}}\right)^{-2}&=&\prod_{a > 0, n}\left(\left(\f{\hat{r}_+ a}{r_0}^2-\omega_i^2\right)^2\right)\prod_{n,N}\left(\left(\f{n}{r_0}+\omega_r \right)^2+\omega_i^2\right)\nonumber\\
&\times& \f{\Gamma\left(1+\left(\f{\omega_i r_0}{\hat{r}_+}\right)\right)^2}{\Gamma\left(1- i\f{r_0}{\hat{r}_+}\left(\f{n}{r_0}+\omega_r\right)+\f{\omega_i r_0}{\hat{r}_+}\right)\Gamma\left(1+i\f{r_0}{\hat{r}_+}\left(\f{n}{r_0}+\omega_r\right)+\f{\omega_i r_0}{\hat{r}_+}\right)}\nonumber\\
&\times& \f{\Gamma\left(1-\left(\f{\omega_i r_0}{\hat{r}_+}\right)\right)^2}{\Gamma\left(1- i\f{r_0}{\hat{r}_+}\left(\f{n}{r_0}+\omega_r\right)-\f{\omega_i r_0}{\hat{r}_+}\right)\Gamma\left(1+i\f{r_0}{\hat{r}_+}\left(\f{n}{r_0}+\omega_r\right)-\f{\omega_i r_0}{\hat{r}_+}\right)}.
\eeqa
This is the most general form of the partition function. We next consider some specific cases.

\newpage

\noindent
\textbf{Case I:} To write the Gamma functions in terms of sine hyperbolic function we require $\f{\omega_i r_0}{\hat{r}_+}=\tilde{N}$ where $\tilde{N}\in\mathbb{Z}$. This gives
\beqa
\left(\f{Z^{(1)}(m)}{e^{\text{Pol}(m)}}\right)^{-2}&=&\prod_{a > 0, n}\left(\left(\f{\hat{r}_+ a}{r_0}^2-\omega_i^2\right)^2\right)\prod_{n,N}\left(\left(\f{n}{r_0}+\omega_r \right)^2+\omega_i^2\right)\nonumber\\
&\times&\f{\Gamma\left(1-\tilde{N}\right)^2\Gamma\left(1+\tilde{N}\right)^2 \sinh\left(\pi\f{r_0}{\hat{r}_+}\left(\f{n}{r_0}+\omega_r\right)\right)^2}{\left(\pi\f{r_0}{\hat{r}_+}\left(\f{n}{r_0}+\omega_r\right)\right)^2 \prod\limits^{\tilde{N}}_{k=1}\left(k^2+\f{r_0^2}{\hat{r}_+^2}\left(\f{n}{r_0}+\omega_r\right)^2\right)}.
\eeqa
At this stage however, the first line does not represent a simple polynomial in $m$ and hence cannot be absorbed in $e^{\text{Pol}(m)}$. We thus consider more specific cases now.

\medskip

\noindent
\textbf{Case II:}
Taking the $m^2\rightarrow \infty$ limit we obtain
\beqa
\omega_r^2 = m^2 +P_n^2-Q_N^2 + \mathcal{O}(m^{-2}) ~~~~ \mbox{and} ~~~~
\omega_i^2 =P_n^2 Q_N^2/m^2 + \mathcal{O}(m^{-4}).
\eeqa
Considering only upto the $\mathcal{O}(m^2)$ terms, we have the following dominant frequencies
\beqa
\omega_r = m ~~~~ \mbox{and} ~~~~
\omega_i = 0.
\eeqa
Next we plug these quasi-normal frequencies in equation (\ref{genZ})
and absorb the $m$ independent terms and the terms that go as simple polynomials in $m$ in $e^{\text{Pol}(m)}$. This gives
\beqa
\left(\f{Z^{(1)}(m)}{e^{\text{Pol}(m)}}\right)^{-1}=\prod_{n}\sinh\left(\f{r_0}{\hat{r}_+}\left(\f{n}{r_0}+m\right)\right)
\eeqa
which can further be written as
\beqa
\left(\f{Z^{(1)}(m)}{e^{\text{Pol}(m)}}\right)^{-1}=\prod_{ n}\left( 1-\exp\left(-\f{2 \pi r_0}{\hat{r}_+}\left(\f{n}{r_0}+m \right)\right)\right).
\eeqa
We can now let $n=k_1-k_2$ to give the partition function to be
\beqa
\f{Z^{(1)}(m)}{e^{\text{Pol}(m)}}=\prod^{\infty}_{k_1,k_2=0}\f{1}{1-e^{\f{2\pi}{\hat{r}_{+}} \left(-m r_0 -(k_1-k_2) \right)}}.
\eeqa
In terms of $\eta$ and $\rho$ this can be written as
\beqa
\f{Z^{(1)}(m)}{e^{\text{Pol}(m)}}=\prod^{\infty}_{k_1,k_2=0}\f{1}{1-e^{\f{2\pi i}{\hat{r}_{+}} \left(\f{\eta}{\hat{r}_+}(k_1-k_2) +i m \rho\right)}}
\eeqa
where $\eta=i \hat{r}_+$ and $\rho=-r_0$. This matches with the result in \cite{Bagchi:2023ilg} up to a factor of $\hat{r}_+$, which can be attributed to a choice of thermal frequencies in the paper. The partition function obtained also matches with \cite{Barnich:2015mui} upon the identification of $(\f{\eta}{\hat{r}_+^2}, \f{\rho}{\hat{r}_+}) \rightarrow \left(\f{\theta}{2 \pi},\f{\beta}{2 \pi}\right)$.

Although the above result and matching to known answers in the literature is very encouraging, we are unable to give any physical justification as to why this particular limit of $m^2\to \infty$ yields this matching. We hope to clarify this rather confusing point in the near future. We make some related remarks in section \ref{conc}.

\bigskip

\noindent
\textbf{Case III:} Emboldened with our previous success, we now explore other regions of parameter space. Let $N=0$ and $m\rightarrow 0$. In this case
\begin{eqnarray}
    \omega_i=0 ~~~\mbox{and}~~~\omega_r=\f{n}{r_0}+\f{m^2 r_0}{2n^2}.
\end{eqnarray}
Thus plugging these equations in equation (\ref{genZ}) we get
\beqa
\f{Z^{(1)}(m)}{e^{\text{Pol}(m)}}=\prod_{ n}\f{1}{ 1-\exp\left(-\f{2 \pi r_0}{\hat{r}_+}\left(\f{2n}{r_0}+\f{m^2 r_0}{2n^2} \right)\right)}.
\eeqa

\noindent
\textbf{Case IV:}
Next, we study the case when $m=0$. Here, our quasi-normal frequencies become
\begin{eqnarray}
    \omega_i=\f{\hat{r}_+ N}{r_0} ~~~\mbox{and}~~~\omega_r=\f{n}{r_0}. 
\end{eqnarray}
Thus, following a similar analysis as before, we get
\beqa
\left(\f{Z^{(1)}(m)}{e^{\text{Pol}(m)}}\right)^{-2}&=&\prod_{a > 0,n,N} \left( \f{i \hat{r}_{+}a}{r_0}-\f{n}{r_0}-\f{n}{r_0}-\f{i \hat{r}_{+}N}{r_0}\right)  \left( \f{i \hat{r}_{+}a}{r_0}-\f{n}{r_0}-\f{n}{r_0}-\f{i \hat{r}_{+}N}{r_0}\right)\nonumber\\
&\times& \prod_{a > 0,n,N} \left( -\f{i \hat{r}_{+}a}{r_0}-\f{n}{r_0}-\f{n}{r_0}+\f{i \hat{r}_{+}N}{r_0} \right) \left( -\f{i \hat{r}_{+}a}{r_0}-\f{n}{r_0}-\f{n}{r_0}+\f{i \hat{r}_{+}N}{r_0}\right)\nonumber\\
&\times& \prod_{n,N} \left(-\f{n}{r_0}-\f{n}{r_0}-\f{i \hat{r}_{+}N}{r_0}\right)\left(-\f{n}{r_0}-\f{n}{r_0}-\f{i \hat{r}_{+}N}{r_0} \right).
\eeqa
This can again be further simplified to further give

\beqa
\left(\f{Z^{(1)}(m)}{e^{\text{Pol}(m)}}\right)^{-2}&=&\left(\prod_{a > 0, n,N}\left[\left(\f{\hat{r}_{+}(a-N)}{r_0}\right)^2+\left(\f{2 n}{r_0}\right)^2\right]\right)^2 \nonumber \\
&\times&\prod_{n,N} \left(\left(\f{\hat{r}_{+}N}{r_0}\right)^2+\left(\f{2 n}{r_0}\right)^2\right).
\eeqa
In this case we regulate the product over $n$ to get the expression
\beqa
\f{Z^{(1)}(m)}{e^{\text{Pol}(m)}}=\prod^{\infty}_{a,N=0}\f{1}{1-\exp\left(- \pi \hat{r}_+\left(a-N\right)\right)}.
\eeqa
We now provide a comparative summary of the quasi-normal frequencies and partition functions that we have obtained in different limits in table \ref{pfnresults}.
\begin{table}[H]
\centering
\renewcommand{\arraystretch}{2.7}
\begin{tabular}{|c|c|c|}
\hline
   \textbf{Conditions} & \textbf{quasi-normal frequencies} & \textbf{One-loop partition function} $ \left(\f{Z^{(1)}(m)}{e^{\text{Pol}(m)}}\right)^{-1}$ \\[10pt]
\hline
  $m\rightarrow \infty$  & $\omega_i=0 , \omega_r=m$ & $\prod\limits^{\infty}_{k_1,k_2=0}\left(1-\exp\left(\f{2\pi }{\hat{r}_{+}} \left((k_1-k_2) + m r_0\right)\right)\right)$ \\[10pt]
\hline
 $N=0, m\rightarrow 0$  & $\omega_i=0 ,\omega_r=\f{n}{r_0}+\f{m^2 r_0}{2n}$ & $ \prod\limits_{ n}\left( 1-\exp\left(-\f{2 \pi r_0}{\hat{r}_+}\left(\f{2n}{r_0}+\f{m^2 r^2_0}{2n^2} \right)\right)\right)$ \\[10pt]
\hline
   $m=0$ &   $\omega_i=\f{\hat{r}_+ N}{r_0},\omega_r=\f{n}{r_0}$  & $\prod\limits^{\infty}_{a,N=0}\left(1-\exp\left(- \pi \hat{r}_+\left(a-N\right)\right)\right)$
  \\[10pt]
\hline
\end{tabular}
 \caption{Partition Functions in various approximations}
\label{pfnresults}
\end{table}

\section{Conclusions and discussions}\label{conc}

\subsection*{Brief Summary}

In this work, we have investigated gravity in three-dimensional asymptotically flat spacetimes. 
The most general zero mode solutions in 3D AFS (with boundary conditions as in \cite{Barnich:2006av}) are flat space cosmologies which are time-dependent solutions with cosmological horizons. In the previous sections, we have studied the properties of the scalar perturbations in this FSC background. We imposed a hard wall boundary at the cosmological horizon and solved the differential equation of the scalar in this background. The solution can also be found by taking the flat space limit from solutions in the BTZ background in AdS$_3$. Finally in this part, we found expressions for quasi-normal modes of the FSC. In the later part of the paper, we used these quasi-normal modes to build the one-loop partition function and in some limit of parameter space, reproduced known results in literature. 

\subsection*{Discussions}

As briefly alluded to in the introduction, duals of 3D AFS have been attempted in terms of a 2D dual Carroll CFT. It turns out that the 2D Conformal Carroll algebra is isomorphic to BMS$_3$ \cite{Bagchi:2010eg, Duval_2014} (the isomorphism holds for higher dimensions as well) and hence Carroll CFTs satisfy the basic requirement of a holographic dual theory to AFS, since the symmetries of the bulk match those of the lower dimensional field theory. 

Following holography of the more studied and more understood AdS spacetimes, specifically AdS$_3$/CFT$_2$ where the BTZ black holes are dual to thermal states in the dual 2D CFT, we expect FSCs to play an analogous role for AFS$_3$/CCFT$_2$. Specifically the expectation is that FSCs should correspond to thermal states in the dual field theory. As mentioned before, this point of view has led to the understanding of the FSC entropy in terms of an underlying counting of states given by a BMS-Cardy formula in the 2D Carroll CFT \cite{Bagchi:2012xr}. There have been several other successes of the AFS$_3$/CCFT$_2$ proposal, including a matching of bulk and boundary correlation functions and entanglement entropy. QNM of the BTZ black hole correspond to linear response and more specifically to the poles in the retarded Green's function. Naively, one would assume a similar relation should also hold in the dual Carroll theory. This is however more subtle. 

Ward identities of Carrollian CFTs in dimensions $d>2$ lead to two branches of correlation functions, viz. the delta-function branch and the CFT branch \cite{Bagchi:2022emh}. E.g. in $d=3$, the two-point correlators of two Carroll primaries with weights $\Delta, \Delta'$ and spins $\sigma, \sigma'$ have the form
\begin{align}
    G^{(2)}_{\delta-fn} &= u_{12}^{2-\Delta - \Delta'} \delta(z_{12}) \delta(\bar{z}_{12}) \delta_{\sigma, \sigma'}; \quad 
    G^{(2)}_{CFT} = z_{12}^{-\Delta - \sigma} \, \bar{z}_{12}^{-\Delta+ \sigma} \, \delta_{\Delta, \Delta'} \delta_{\sigma, \sigma'}. 
\end{align}
In $d=2$, the zero modes $L_0$ and $M_0$ commute and hence the states labeled by eigenvalues of $L_0$ also need to be labelled by $M_0$. $M_0$ also turns out to be non-diagonal in the highest weight representations. The structure of correlations are thus more complicated and although there is quite a bit of work on the ``CFT'' branch, the delta-function branch in $d=2$ has not been addressed in detail in the literature so far. Thermal Green's functions are also expected to have two branches and hence the pole structure and possibly the matching of FSC QNM to a putative field theory analysis would depend on these details. A quick analysis did not yield satisfactory results and we aim to understand linear response and thermal Green's functions in Carrollian QFTs and CFTs. Keeping in mind the subtelties of different branches of correlation functions, we would return to the matching of the poles of these correlations to the spectrum of QNM we have uncovered in this work. 

We conclude with another intriguing point. The construction of the one-loop partition functions using the DHS method with the spectrum of QNM that we discovered for the FSC yielded a matching with known answers only in a certain limiting case ($m\to\infty$). The previous answers in the literature also reproduced the character of the underlying BMS$_3$ algebra. The fact that we seem to have a more general answer for the 1-loop partition function of 3D AFS is interesting and could be indicating several things. It may be possible that the boundary conditions that we implicitly use for the scalar field in the FSC background are more general than the Barnich-Compère boundary conditions and there are enhancements in the asymptotic symmetry algebra which reduces to the BMS only in a certain limit (see e.g. \cite{Grumiller:2017sjh}) {\footnote{This is also reminiscent of the $w_{1+\infty}$ enhancements of the BMS$_4$ in $D=4$ AFS \cite{Strominger:2021mtt}.}}. It is clear that the analysis of \cite{Barnich:2006av} was tailor made for metrics of the form \eqref{BC-metric}, but our analysis has been performed in the changed coordinates \eqref{met}. A concrete way to understand the issue would be to redo the canonical analysis in \eqref{met} with appropriate boundary conditions and this may lead to a different asymptotic symmetry algebra and hence a different candidate dual theory. Perhaps the characters of these more general algebras are key to our analyses. In any case, it would be very instructive to figure out what the QNMs of the FSC that we have uncovered indicate in terms of connections to 1-loop partition functions of 3D AFS. We intend to return to this in the near future.

\bigskip

\subsection*{Acknowledgements}
We thank Stéphane Detournay and Daniel Grumiller for interesting discussions. 

\smallskip

\noindent AB is partially supported by a Swarnajayanti Fellowship from the Science and Engineering Research Board
(SERB) under grant SB/SJF/2019-20/08 and also by an ANRF grant CRG/2022/006165. 

\smallskip

\noindent SB is supported by an IIT Kanpur Institute Assistantship. 

\smallskip

\noindent AK acknowledges the support provided by SB/SJF/2019-20/08. AK would also like to acknowledge the organizers of Holostrings II workshop for interesting discussions and thank Chennai Mathematical Institute, CERN, Durham University and University of Liverpool for their warm hospitality and interesting discussions during the course of this work. 

\smallskip

\noindent SM is supported by grant number 09/092(1039)/2019-EMR-I from Council of Scientific and Industrial Research (CSIR). SM would also like to acknowledge the hospitality of Erwin Schrödinger International Institute for Mathematics and Physics, University of Vienna and Institute for Theoretical Physics, TU Wien, where a part of the work was carried out.

\section*{APPENDICES}
\begin{appendix}
\section{Further details of scalar field in FSC}
\subsection{Ingoing and outgoing boundary condition for FSCs}\label{inout}
We check for the ingoing and outgoing condition at the horizon as done for AdS blackholes and dS space respectively. Near the horizon $\tau\to0$, the solutions behave as follows:
\begin{equation}
\psi^{I,\pm}_{n,p}=\frac{\left(\frac{\omega\tau}{2}\right)^{\pm i\frac{p}{E}}}{\Gamma(1\pm ip/E)}e^{ipx}e^{i\frac{n}{r_0}y}=\frac{e^{ip\left(x\pm\frac{\tau_*}{E}\right)}e^{\pm i\frac{\ln(\omega/2)}{E}}}{\Gamma(1\pm ip/E)}e^{i\frac{n}{r_0}y}
\end{equation}
$\psi^+$ and $\psi^-$ are respectively ingoing and outgoing at the horizon. To make the ingoing solution vanishing at the horizon, we make $\Gamma(1+i\frac{p}{E})=\infty$, which yields $p=iEN,\ N\in\mathbb{N}$. Under this condition, $\psi^-$ takes the following form
\begin{equation}
\psi^{I,-}_{n,N}=\frac{\left(\frac{\omega\tau}{2}\right)^{N}}{\Gamma(1+N)}e^{-ENx}e^{i\frac{n}{r_0}y}.
\end{equation}
This is well-behaved for $x>0$ but diverges for $x\to-\infty$.
\par Now, to make the outgoing solution vanish near the horizon, we make $\Gamma(1-i\frac{p}{E})=\infty$, which yields $p=-iEN,\ N\in\mathbb{N}$. Under this condition, $\psi^+$ takes the following form
\begin{equation}
\psi^{I,+}_{n,N}=\frac{\left(\frac{\omega\tau}{2}\right)^{N}}{\Gamma(1+N)}e^{ENx}e^{i\frac{n}{r_0}y}.
\end{equation}
This is well-behaved for $x<0$ but diverges for $x\to\infty$. Thus, we see the only consistent way is to make both the solutions vanish near the horizon (which means reflective boundary condition).

\subsection{Scalar fields in the inner region of FSC} \label{innerfsc}
Consider the following differential equation satisfied by the scalar fields in the inner region of FSC $(r < r_0)$,
\begin{equation}\label{innbess}
    \left[\rho^2\frac{d^2}{d\rho^2}+\rho\frac{d}{d\rho}+((i\omega)^2\rho^2-\nu^2)\right]f^{II}(\rho)=0 
\end{equation}
where $\omega^2=(p-n/r_0)^2+m^2$ and $\nu=i\frac{p}{E}$. The solutions to this equation are of the form  
\begin{equation}
    \psi_{n,p}^{II,\pm}=J_{\pm\nu}(i\omega|\rho|)e^{ipx}e^{i\frac{n}{r_0}y}\,.
\end{equation}
For non-integer $\nu$, $\psi^{\pm}$ are two independent solutions. 

Now, in this region, $x$ is timelike. We impose the standard boundary condition keeping in mind that $r=r_0$ is a cosmological horizon,
\begin{enumerate}
    \item The solutions are regular inside the region $r<r_0$.
    \item The solutions are purely outgoing at the cosmological horizon $\rho \rightarrow 0 $. 
\end{enumerate}
As $\rho \to 0$, the solutions behave as 
\begin{equation}
    \psi_{n,p}^{II,\pm} \sim e^{ipx}e^{i\frac{n}{r_0}y}\frac{1}{\Gamma(1\pm i\frac{p}{E})}\left(\frac{i\omega\rho}{2}\right)^{\pm i\frac{p}{E}}\,.
\end{equation}

Now, since we do not need to worry about the behavior as $x\rightarrow \infty$, the fate of ingoing and outgoing modes are not intertwined. We thus may impose the outgoing boundary condition at the cosmological horizon in this region.

\section{Scalar field in AdS}\label{adsqnm}
\subsection{Scalar field solutions}
We now discuss the quasi-normal frequencies in AdS following the analysis in \cite{Aharony:1999ti}. The AdS metric in global coordinates is given by
\beqa
ds^2=\f{\ell^2}{\cos^2 \theta}(-d\tau^2+d\theta^2+\sin^2\theta d\Omega^2)\,,
\eeqa
where $\ell$ is the AdS radius of curvature. We consider a massive scalar field in AdS$_{d+1}$ which satisfies the following equation
\beqa
(\Box-m^2)\phi=0.
\eeqa
This equation is solved by the following wave functions
\beqa \label{adssol}
\phi=e^{i\omega \tau}R(\theta)Y_l(\Omega_d)\,,
\eeqa
where $Y_l(\Omega_d)$ are the spherical harmonics that emerge as the eigenstates of the Laplacian on $S^{d-1}$ with eigenvalues $l(l+d-1)$ and $R(\theta)$ is the radial function given by
\beqa \label{adsradial}
R(\theta)&=&\mathcal{C}(\sin \theta)^l (\cos \theta)^{\lambda_{\pm}}~_2F_1(a,b,c;\sin \theta)\nonumber\\
&+&\mathcal{D}(\sin \theta)^{1-l} (\cos \theta)^{\lambda_{\pm}}~_2F_1(a-c+1,b-c+1,2-c;\sin \theta)
\eeqa
where we have defined the following
\beqa
a&=&\f{1}{2}(l+\lambda_{\pm}-\omega \ell) , \nonumber \\
b&=&\f{1}{2}(l+\lambda_{\pm}+\omega \ell) , \nonumber \\
c&=&l+\f{1}{2}(d+1), \nonumber \\
\lambda_{\pm}&=&\f{1}{2}(d+1)\pm\f{1}{2}\sqrt{(d+1)^2+4(m \ell)^2}.
\eeqa
and $\mathcal{C}$ and $\mathcal{D}$ are some constants. The anti-de Sitter QNMs are solutions to the equations of motion
that satisfy the physical boundary conditions: (i) the field is regular at the origin; (ii) the
flux is vanishing at the asymptotic boundary. Since we need our solutions to be regular at $\theta=0$, we need to set $\mathcal{D}=0$ as the second term in equation \ref{adsradial} has a singularity there.
As we are aware, massive particles can never get to the boundary of AdS. This can be seen using the reflective boundary conditions at the boundary or imposing the requirement that the energy-momentum flux through the boundary at $\theta=\pi/2$ vanishes. In general, the energy-momentum tensor, with some general $\beta$ which depends on the coupling of the scalar curvature to $\phi^2$, is conserved and is given by
\beqa
T_{\mu \nu}=2\partial_\mu \phi \partial_\nu \phi -g_{\mu\nu}\left((\partial\phi)^2+m^2\phi^2\right)+\beta(g_{\mu\nu}\Delta-D_{\mu}D_{\nu}+R_{\mu\nu})\phi^2\,.
\eeqa
Now, we notice that the total energy $E$ of the scalar field fluctuation, $E=\int d^{d+1} x \sqrt{-g}T_0^0$, is conserved only if the energy-momentum flux through the boundary vanishes. We see this as follows
\beqa
&&\int_{S^{d-1}}d\Omega_{d-1}\sqrt{g}n_i T^i_{0}\big|_{\theta=\pi/2}=0\nonumber\\
&&\rightarrow(\tan \theta)^d[(1-2\beta)\partial_{\theta}+2\beta\tan\theta]\phi^2\big|_{\theta=\pi/2}\rightarrow 0\,.
\eeqa
For us $\beta=0$ since we are not considering any coupling to the curvature. Plugging in the scalar field solution from equation \ref{adssol} we get the requirement that either $a$ or $b$ should be an integer. This gives us the following quasi-normal frequencies for AdS
\beqa\label{adszf}
\omega_{n,l}=\pm\f{\lambda_{\pm}+l+2n}{\ell} ~~~~~~~~~(n=0,1,2,\cdots)
\eeqa
where $\omega$ is real.

\subsection{One-loop partition function}
Thermal frequencies are $\omega_n=2\pi n T$ where $T$ is the temperature derived from the periodicity in the Euclidean time direction.
Using DHS prescription, which has already been discussed in detail in the previous sections, the AdS one-loop partition function can be written as
\beqa \label{adspf}
Z^{(1)}=e^{\text{Pol}(\Delta)}\prod_{z_{\star}\overline{z}_{\star}}\f{\sqrt{z_{\star}\overline{z}_{\star}}}{4\pi^2 T}\Gamma\left(\f{iz_{\star}}{2\pi T}\right)^{-1}\Gamma\left(-\f{i\overline{z}_{\star}}{2\pi T}\right)^{-1}=e^{\text{Pol}(\Delta)}\prod_{z_{\star}}\f{e^{-\f{|z_{\star}|}{2 T}
}}{1-e^{-\f{|z_{\star}|}{T}}}\,.
\eeqa
Sticking to the Dirichlet boundary conditions we keep only $\lambda_{+}$ and replace it by the usual notation $\Delta$. We also need to take care of the degeneracy $D_l$ of each frequency which is basically the degeneracy of the $l^{th}$ angular momentum eigenvalue on $S^{d-1}$. We now specialize to the case of a real scalar in AdS$_3$ and thus need to take the square root of equation \ref{adspf}. Substituting the QNMs from \ref{adszf} we get
\beqa
Z^{(1)}=e^{\text{Pol}(\Delta)-\sum_{l,n}2D_l\f{2n+l+\Delta}{2 \ell T}}\prod_{n,l}\left(1-e^{-\f{2n+l+\Delta}{\ell T}}\right)^{-2D_l}\,.
\eeqa

We also combine the products over $l,n$ and the multiplicity into a single product over $k=2n+l$ with multiplicity $k+1$ and after absorbing UV divergences depending on $\Delta$ polynomially into $\text{Pol}(\Delta)$ to again get
\beqa
Z^{(1)}=e^{\text{Pol}(\Delta)}\prod_{l,l'= 0}\f{1}{\left(1-q^{l+\Delta/2}\bar{q}^{l'+\Delta/2} \right)}
\eeqa
where $q=\overline{q}=e^{-2\pi/T}$ now in this case.

\section{Scalar fields in BTZ black hole }\label{APPA}
\subsection{Scalar fields in the outer region of BTZ black hole }
In this section, we will briefly review scalar fields in the BTZ background \cite{Birmingham_chptuik, Birmingham:2001pj} and then perform a flat space limit over the computation. Since the FSC can also be obtained by taking the flat-space limit ($\ell\to\infty$) from the BTZ black hole, we are inspired to study this limit. There, however, are certain peculiarities and differences that we observe along the way.

The metric of a non-extremal BTZ black hole is given by
\begin{equation}\label{btz2}
\begin{aligned}
    ds^2=-\frac{(r^2-r_+^2)(r^2-r_-^2)}{r^2\ell^2}dt^2+\frac{\ell^2r^2dr^2}{(r^2-r_+^2)(r^2-r_-^2)}
    +r^2\left(d\varphi-\frac{r_+r_-}{r^2\ell}dt\right)^2.
\end{aligned}
\end{equation}
Under the following parametrization in the region $(r>r_+)$
\begin{eqnarray}
r^2&=&\frac{r_+^2-r_-^2}{2}\cosh2\mu+\frac{r_+^2+r_-^2}{2},\nonumber\\
x^+&=&\frac{r_+}{\ell}t-r_-\varphi,\ x^-=r_+\varphi-\frac{r_-}{\ell}t,\nonumber\\
z&=&\tanh^2\mu=\frac{r^2-r_+^2}{r^2-r_-^2},
\end{eqnarray}
the BTZ metric takes the following form
\begin{align}
ds^2&=-\sinh^2\mu(dx^+)^2+\ell^2d\mu^2+\cosh^2\mu(dx^-)^2,\\
ds^2&=-\frac{z}{1-z}(dx^+)^2+\frac{\ell^2dz^2}{4z(1-z)^2}+\frac{(dx^-)^2}{1-z}.\label{abc}
\end{align}
The scalar wave equation that one needs to solve is 
\begin{equation}
    (\Box-m^2)\Psi=0\,.
\end{equation}
Substituting the metric \eqref{abc} in the wave equation and using the ansatz
\begin{equation}
\psi=e^{-i(k_+x^++k_-x^-)}R(z),
\end{equation}
the Klein-Gordon equation yields
\begin{equation}\label{kgeq}
z(1-z)\frac{d^2R}{dz^2}+(1-z)\frac{dR}{dz}+\left[\frac{k_+^2\ell^2}{4z}-\frac{k_-^2\ell^2}{4}-\frac{m^2\ell^2}{4(1-z)}\right]R=0.
\end{equation}
We further assume that
\begin{eqnarray}
R(z)=z^\alpha(1-z)^\beta F(z).
\end{eqnarray}
Under this assumption, equation (\ref{kgeq}) reduces to a hypergeometric ODE of the following form
\begin{equation}
z(1-z)\frac{d^2F}{dz^2}+[c-(1+a+b)z]\frac{dF}{dz}-abF=0,
\end{equation}
where we have defined

\beqa
&&\displaystyle \alpha=-\frac{ik_+\ell}{2},\nonumber\\
&& \beta=1-\frac{\Delta}{2},\ c=1-ik_+\ell,\nonumber\\
&&  \displaystyle a=1-\frac{\Delta}{2}-\frac{i\ell}{2}(k_+-k_-),\nonumber\\
&&   b=1-\frac{\Delta}{2}-\frac{i\ell}{2}(k_++k_-).
\eeqa
We have also used the fact that for a relativistic scalar field theory in 3D, the scaling dimension is $\Delta=1+\sqrt{1+m^2\ell^2}$. Around $z=0$ the hypergeometric ODE has two independent solutions, ${}_2F_1(a,b;c;z)$ and $z^{1-c}{}_2F_1(a-c+1,b-c+1;2-c;z)$. Thus, in our case as well we have two independent solutions that are given  by
\begin{align}
R_1(z)&=z^{-\frac{ik_+\ell}{2}}(1-z)^{1-\frac{\Delta}{2}}{}_2F_1(a,b;c;z),\\
R_2(z)&=z^{\frac{ik_+\ell}{2}}(1-z)^{1-\frac{\Delta}{2}}{}_2F_1(a-c+1,b-c+1;2-c;z).
\end{align}

\begin{figure}
\centering
  \includegraphics[width=1.0\linewidth]{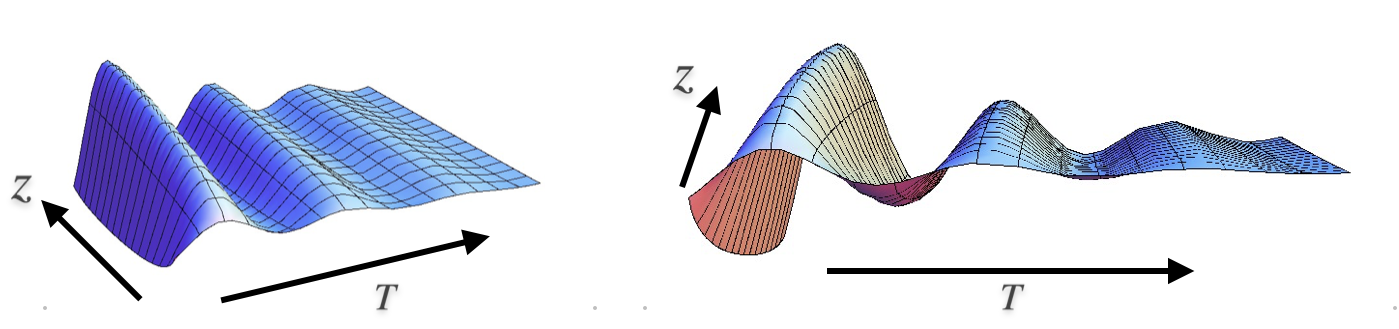}
  \label{fig:sub21}

\caption{Representative incoming solution for BTZ.}
\label{f(x)solutionBTZ}
\end{figure}

The flux in this case is defined as
\begin{equation}
\mathcal{F}:=\frac{\sqrt{-g}}{\ell}\text{Im}(R^*\partial_\mu R)=\frac{\sqrt{z}}{1-z}\text{Im}(R^*\partial_z R).
\end{equation}
Near the horizon the flux becomes
\begin{equation}
    \lim_{z\to0^+}\mathcal{F}_1=\left(-\frac{k_{+,r}\ell}{2\sqrt{z}}+\Im\left(\frac{ab}{c}\right)\sqrt{z}\right)z^{\ell k_{+,i}}.
\end{equation}
We note that if $k_{+,r}>0$, then this flux $\mathcal{F}_1<0$, \textit{i.e.} $R_1(z)$ is the ingoing wave at the horizon, and additionally, if $k_{+, i}>\frac{1}{2}$, the wave function is regular at the horizon. Under these conditions, $R_2(z)$ has to be the outgoing wave solution. A quick check reveals that
\begin{equation}
    \lim_{z\to0^+}\mathcal{F}_2=\left(\frac{k_{+,r}\ell}{2\sqrt{z}}+\Im\left(\frac{(a-c+1)(b-c+1)}{2-c}\right)\sqrt{z}\right)z^{-\ell k_{+,i}}>0\,.
\end{equation}
The BTZ QNMs are solutions to the equations of motion
that satisfy the physical boundary conditions: (i) the field is ingoing at the event horizon; (ii) vanishing at infinity. Therefore only $R_1(z)$, which is ingoing at the horizon, survives.
We further observe that at spatial infinity ($z\to1$), the flux  must be zero. To see this we let $z=1-\epsilon$, where $\epsilon>0$ is an infinitesimally small parameter. The flux then becomes
\begin{equation}
    \lim_{z\to1}\mathcal{F}_1=-\frac{\ell k_{+,r}}{2\epsilon^{(\Delta-1)}}\left|\frac{\Gamma(c)\Gamma(c-a-b)}{\Gamma(c-a)\Gamma(c-b)}\right|^2+\frac{1}{\epsilon^{(\Delta-2)}}\Im\left(\frac{ab}{c-a-b-1}\right)\left|\frac{\Gamma(c)\Gamma(c-a-b)}{\Gamma(c-a)\Gamma(c-b)}\right|^2\,.
\end{equation}
Hence, we see that the flux is scaled with $\displaystyle\left|\frac{\Gamma(c)\Gamma(c-a-b)}{\Gamma(c-a)\Gamma(c-b)}\right|^2$. In order to make the flux vanishing, the denominator must diverge, \textit{i.e.} we require that $c-a=-N,\ \text{or},\ c-b=-N$. This leads us to conclude that
\begin{equation}
k_+\pm k_-=-i\frac{2N+\Delta}{\ell}
\end{equation}
and $\displaystyle k_{+,r}=\mp k_{-,r},\ \ell k_{+,i}=-(2N+\Delta)\mp\ell k_{-,i}$. Hence, $k_{-,r}<0$. If we now write the phase using another set of coordinates as $k_+x^++k_-x^-=\omega t-k\varphi$, we obtain the relations
\begin{eqnarray}
k_++k_-=\frac{\omega\ell-k}{r_+-r_-},\quad k_+-k_-=\frac{\omega\ell+k}{r_++r_-}.
\end{eqnarray}
Here, we have defined 
\begin{equation}
T_L=\frac{r_+-r_-}{2\pi\ell^2};\quad T_R=\frac{r_++r_-}{2\pi\ell^2}
\end{equation}
as the temperatures of two dual conformal field theory sectors at thermal equilibrium \cite{Birmingham_chptuik, Birmingham:2001pj, Birmingham_2003}. One might recall that according to $\text{AdS}_{3}/\text{CFT}_2$ correspondence, each spin $s$ propagating in $\text{AdS}_{3}$ corresponds to an operator $\mathcal{O}$ in $\text{CFT}_2$ characterised by conformal weights ($h_L,\ h_R$) with $h_R+h_L=\Delta$ and $h_R-h_L=s$. Here, in this case, $s=0$ and $h_L=h_R=\Delta/2$. Thus, we get the following QNMs for the BTZ black hole
\begin{eqnarray}
\omega_L&=\frac{k}{\ell}-4\pi iT_L(N+h_L)\nonumber\\
\omega_R&=-\frac{k}{\ell}-4\pi iT_R(N+h_R)\,.
\end{eqnarray}
We observe that the frequencies are complex, which confirms that the system is open and dissipating energy. Also, it is seen that the imaginary part scales linearly with the temperature.
\subsection{Scalar fields in the intermediate region of BTZ black hole} 
In the \textit{intermediate region} ($r_-<r<r_+$), $r$ becomes timelike. The metric is 
\begin{equation}
\begin{aligned}
    ds^2=\frac{(r_+^2-r^2)(r^2-r_-^2)}{r^2\ell^2}dt^2-\frac{\ell^2r^2dr^2}{(r_+^2-r^2)(r^2-r_-^2)}
    +r^2\left(d\varphi-\frac{r_+r_-}{r^2\ell}dt\right)^2\,.
\end{aligned}
\end{equation}
In our region of interest, we choose the two following coordinate parametrization in terms of $(\tilde{\mu},x^+,x^-)$ and $(z,x^+,x^-)$ coordinates
\begin{eqnarray}
    x^+=\frac{r_+}{\ell}t-r_-\varphi~,~\ x^-=r_+\varphi-\frac{r_-}{\ell}t~,~\frac{r^2-r_-^2}{r_+^2-r_-^2}=\sin^2\tilde{\mu}=z\,.
\end{eqnarray}
The metrics in these coordinates are respectively 
\begin{align}
ds^2&=\cos^2\tilde{\mu}(dx^+)^2-\ell^2d\tilde{\mu}^2+\sin^2\tilde{\mu}(dx^-)^2,\\
ds^2&=(1-z)(dx^+)^2-\frac{\ell^2dz^2}{4z(1-z)}+z(dx^-)^2.
\end{align}
We study a complex, massive scalar field in this background. Under the minimal coupling condition, substituting the metric in the wave equation, we get,
\begin{equation}
    z(1-z)\partial^2_z\psi+(1-2z)\partial_z\psi-\frac{\ell^2}{4}\left[\frac{\partial^2_+\psi}{1-z}+\frac{\partial^2_-\psi}{z}\right]+\frac{m^2\ell^2}{4}\psi=0\,.
\end{equation}
To solve the above equation, we take the ansatz
\begin{equation}\label{ansatz}
    \psi=e^{-i(k_+x^++k_-x^-)}R(z)\,,
\end{equation}
which yields a hypergeometric equation
\begin{equation}
    z(1-z)\frac{d^2R}{dz^2}+(1-2z)\frac{dR}{dz}+\left[\frac{\ell^2k_+^2}{4(1-z)}+\frac{\ell^2k_-^2}{4z}+\frac{m^2\ell^2}{4}\right]R(z)=0.
\end{equation}
We can show more manifestly that the above is a hypergeometric equation if we choose the following form of the solution
\begin{equation}
R(z)=z^{\alpha}(1-z)^{\beta}F(z)\,.
\end{equation}
This gives us
\begin{eqnarray}
&z(1-z)\frac{d^2F}{dz^2}+[(2\alpha+1)-(2\alpha+2\beta+2)z]\frac{dF}{dz}-\left[(\alpha+\beta)(\alpha+\beta+1)-\frac{m^2\ell^2}{4}\right]F(z)\nonumber\\
&~~~~~~~~~~~~~~~~~~~~~~~~~~~~~~~~~~~~~~~~~~~~~~~~~+\left[\frac{\beta^2+(k_+\ell/2)^2}{1-z}+\frac{\alpha^2+(k_-\ell/2)^2}{z}\right]F(z)=0.
\end{eqnarray}
We map it to the standard form of the hypergeometric equation
\begin{equation}
    z(1-z)\frac{d^2F}{dz^2}+[c-(1+a+b)z]\frac{dF}{dz}-abF=0
\end{equation}
by identifying the constant parameters as the following
\begin{eqnarray}
    &\alpha=-ik_-\ell/2,\quad\beta=-ik_+\ell/2,\quad c=1-ik_-\ell,\nonumber\\
    &ab=-\frac{\ell^2}{4}[(k_++k_-)^2+m^2]-i\ell(k_++k_-)/2,\nonumber\\
    &a+b=1-i(k_++k_-)\ell,\quad a-b=\sqrt{1+m^2\ell^2}.
\end{eqnarray}
For scalar fields, the scaling dimension is $\Delta=1+\sqrt{1+m^2\ell^2}$ and solving the above one gets,
\begin{eqnarray}
a=\frac{\Delta}{2}-\frac{i\ell}{2}(k_++k_-),\quad b=1-\frac{\Delta}{2}-\frac{i\ell}{2}(k_++k_-).
\end{eqnarray}
The equation has three singular points $0, 1,\infty$. Around each point, we can find two independent solutions as described below.
\begin{itemize}
\item Around $z=0$,
    \begin{eqnarray}
    R(z)&=&A_0z^{-\frac{ik_-\ell}{2}}(1-z)^{-\frac{ik_+\ell}{2}}{}_2F_1(a,b;c;z)\nonumber\\
    &+&B_0z^{\frac{ik_-\ell}{2}}(1-z)^{-\frac{ik_+\ell}{2}}{}_2F_1(1+a-c,1+b-c;2-c;z).
    \end{eqnarray}
\item Around $z=1$,
    \begin{eqnarray}
    R(z)&=&A_1 z^{-\frac{ik_-\ell}{2}}(1-z)^{-\frac{ik_+\ell}{2}}{}_2F_1(a,b;1+a+b-c;1-z)\nonumber\\
    &+&B_1 z^{-\frac{ik_-\ell}{2}}(1-z)^{\frac{ik_+\ell}{2}}{}_2F_1(c-a,c-b;1+c-a-b;1-z).
   \end{eqnarray}
\item Around $z=\infty$,
\begin{eqnarray}
    R(z)&=&A_{\infty}z^{\frac{ik_+\ell}{2}}z^{-\frac{\Delta}{2}}(1-z)^{-\frac{ik_+\ell}{2}}{}_2F_1(a,1+a-c;1+a-b;z^{-1})\nonumber\\
    &+&B_{\infty}z^{\frac{ik_+\ell}{2}}z^{\frac{\Delta}{2}-1}(1-z)^{-\frac{ik_+\ell}{2}}{}_2F_1(b,1+b-c;1+b-a;z^{-1}).
\end{eqnarray}
\end{itemize}
Now, the flux is defined as
\begin{equation}
\mathcal{F}:=-\frac{\sqrt{-g}}{\ell}\text{Im}(R^*\partial_{\tilde{\mu}} R)=-\sqrt{z(1-z)}\text{Im}(R^*\partial_z R).
\end{equation}
The minus sign appears because $\tilde{\mu}$ is a timelike coordinate. For our present purpose, we choose
\begin{eqnarray}
R_1(z)&=&A_0z^{-\frac{ik_-\ell}{2}}(1-z)^{-\frac{ik_+\ell}{2}}{}_2F_1(a,b;c;z),\nonumber\\
R_2(z)&=&B_0z^{\frac{ik_-\ell}{2}}(1-z)^{-\frac{ik_+\ell}{2}}{}_2F_1(1+a-c,1+b-c;2-c;z).
\end{eqnarray}
Close to the inner horizon ($z\to0^+)$ we have the following behavior of the solutions
\begin{eqnarray}
R_1(z)&\sim A_0z^{-\frac{ik_{-,r}\ell}{2}}z^{\frac{k_{-,i}\ell}{2}}~~,~~
R_2(z)&\sim B_0z^{\frac{ik_{-,r}\ell}{2}}z^{-\frac{k_{-,i}\ell}{2}}.
\end{eqnarray}
For $k_{-,i}<0$, $R_1(z)$ diverges and for $k_{-,i}>0$, $R_2(z)$ diverges. We, therefore, need to choose certain values of $A_0$ and $B_0$ to avoid the divergence. We choose
\begin{equation}
    A_0=\frac{1}{\Gamma(1-ik_-\ell)}\ \text{and}\ B_0=\frac{1}{\Gamma(1+ik_-\ell)}.
\end{equation}
So, for $k_{-,i}<0$
\begin{eqnarray}
R_1(z)&\sim z^{-\frac{ik_{-,r}\ell}{2}}\frac{1}{z^{\frac{|k_{-,i}|\ell}{2}}\Gamma(1-|k_{-,i}|\ell-ik_{-,r}\ell)},\nonumber\\
R_2(z)&\sim \frac{1}{\Gamma(1+|k_{-,i}|\ell+ik_{-,r}\ell)}z^{\frac{ik_{-,r}\ell}{2}}z^{\frac{|k_{-,i}|\ell}{2}}.
\end{eqnarray}
We need $\Gamma(1-|k_{-,i}|\ell-ik_{-,r}\ell)$ to diverge which can happen \textit{iff} $|k_{-,i}|\ell=N,\ N\in\mathbb{N}$ and $k_{-,r}\ell=0$.
\par Similarly for $k_{-,i}>0$, we have
\begin{eqnarray}
 R_1(z)&\sim \frac{1}{\Gamma(1+k_{-,i}\ell-ik_{-,r}\ell)}z^{-\frac{ik_{-,r}\ell}{2}}z^{\frac{k_{-,i}\ell}{2}},\nonumber\\
 R_2(z)&\sim z^{\frac{ik_{-,r}\ell}{2}}\frac{1}{z^{\frac{|k_{-,i}|\ell}{2}}\Gamma(1-k_{-,i}\ell+ik_{-,r}\ell)}.
\end{eqnarray}
We now need $\Gamma(1-k_{-,i}\ell+ik_{-,r}\ell)$ to diverge which can happen \textit{iff} $k_{-,i}\ell=N,\ N\in\mathbb{N}$ and $k_{-,r}\ell=0$. So, we can say $k_-\ell=iN,\ N\in\mathbb{Z}$. Thus we have
\begin{eqnarray}
R_1(z)&=z^{-\frac{ik_-\ell}{2}}(1-z)^{-\frac{ik_+\ell}{2}}\sum_{s=0}^{\infty}\frac{(a)_s(b)_s}{\Gamma(s+1-ik_-\ell)}\frac{z^s}{s!}\nonumber\\
&=z^{\frac{N}{2}}(1-z)^{-\frac{ik_+\ell}{2}}\sum_{s=0}^{\infty}\frac{(a)_s(b)_s}{\Gamma(s+1+N)}\frac{z^s}{s!}
\end{eqnarray}
\begin{equation}
R_1(z)=\begin{cases}
&\frac{z^{\frac{N}{2}}(1-z)^{-\frac{ik_+\ell}{2}}}{N!}{}_2F_1(a,b;1+N;z)\quad N>0\\
&z^{\frac{|N|}{2}}(1-z)^{-\frac{ik_+\ell}{2}}\frac{(a)_{|N|}(b)_{|N|}}{|N|!}{}_2F_1(a-N,b-N;1-N;z)\quad N<0
\end{cases}
\end{equation}
and
\begin{eqnarray}
R_2(z)&=&z^{\frac{ik_-\ell}{2}}(1-z)^{-\frac{ik_+\ell}{2}}\sum_{s=0}^{\infty}\frac{(a+ik_-\ell)_s(b+ik_-\ell)_s}{\Gamma(s+1+ik_-\ell)}\frac{z^s}{s!}\nonumber\\
&=&z^{-\frac{N}{2}}(1-z)^{-\frac{ik_+\ell}{2}}\sum_{s=0}^{\infty}\frac{(a-N)_s(b-N)_s}{\Gamma(s+1-N)}\frac{z^s}{s!}
\end{eqnarray}
\begin{equation}
R_2(z)=\begin{cases}
&z^{\frac{N}{2}}(1-z)^{-\frac{ik_+\ell}{2}}\frac{(a-N)_{N}(b-N)_{N}}{N!}{}_2F_1(a,b;1+N;z)\quad N>0\\
&\frac{z^{\frac{|N|}{2}}(1-z)^{-\frac{ik_+\ell}{2}}}{|N|!}{}_2F_1(a-N,b-N;1-N;z)\quad N<0\,.
\end{cases}
\end{equation}
Thus, equivalently we can write two independent solutions with $N\in\mathbb{N}$
\begin{align}
R_-(z;\,N)&=z^{\frac{N}{2}}(1-z)^{-\frac{ik_+\ell}{2}}\frac{{}_2F_1(a,b;1+N;z)}{N!}\\
R_+(z;\,N)&=z^{\frac{N}{2}}(1-z)^{-\frac{ik_+\ell}{2}}\frac{{}_2F_1(a+N,b+N;1+N;z)}{N!}\,.
\end{align}
To see the implications of such a choice, we investigate the behavior of the whole solution. Our ansatz solutions (\ref{ansatz}) are
\begin{align}
\psi^-_N&=e^{-ik_+x^+}\,e^{Nx^-/\ell}\,R_-(z;\,N),\\
\psi^+_N&=e^{-ik_+x^+}\,e^{-Nx^-/\ell}\,R_+(z;\,N).
\end{align}
We demand the solutions to be well behaved at $|x^-|\to\infty$. As a result, we are led to the conclusion that $\psi^-_N$ is defined for $x^-<0$ and $\psi^+_N$ is defined for $x^->0$. We can write any of the solutions as
\begin{equation}
\psi^{\pm}_{{}_{N}}=e^{-N|x^-|/\ell}\,z^{\frac{N}{2}}\ \exp\left[ik_+\left\lbrace\frac{\ell}{2}\ln\left(\frac{1}{1-z}\right)-x^+\right\rbrace\right]\times
\begin{cases}
\frac{{}_2F_1(a,b;1+N;z)}{N!}& \text{or}\\    
\frac{{}_2F_1(a+N,b+N;1+N;z)}{N!}.
\end{cases}
\end{equation}

\subsection{One-loop partition function of scalar field in BTZ} \label{APPB}
As was shown in \cite{Denef:2009kn}, \cite{Denef:2009yy}, the one-loop partition function for a massive scalar field $\phi$ on a thermal AdS$^{}_{d+1}$ background can be expressed in terms of the quasi-normal modes $\omega^{}_{*}(\Delta)$ as

\begin{equation}
\label{enm1}
Z^{(1)}(\Delta)=e^{\text{Pol}(\Delta)}\prod^{}_{a,\star}(\omega^{}_{a}(T)-\omega^{}_{\star}(\Delta))^{-1}.
\end{equation}

In the above equation, $\star$ represents some set of additional quantum numbers apart from the thermal quantum number and $\omega^{}_{a}$ represents the Matsubara frequencies, at temperature $T$, of the thermal background. We discuss how the DHS prescription applies to the BTZ solution.

\vspace{0.2cm}
\noindent

Thermal frequencies are defined as
\beqa
\omega_a =\f{4 \pi i T_L T_R}{T_L+T_R}a+\f{T_R-T_L}{T_L+T_R}k
\eeqa
where $a$ represents the thermal quantum number. As seen from our previous analysis, we get two sets of quasi-normal frequencies, corresponding to the in-going and out-going modes which are either prograde or retrograde
\beqa
\omega_{in}=\pm k- 2\pi i T_R(2 N+\Delta)
\eeqa
and
\beqa
\omega_{out}=\pm k+ 2\pi i T_L(2 N+\Delta).
\eeqa
We can thus write the partition function using \eqref{enm1} as
\beqa
\left(\f{Z^{(1)}(\Delta)}{e^{\text{Pol}(\Delta)}}\right)^{-2}&=&\prod_{a > 0,N\geq 0,k} \left(\omega_a +k+2\pi i T_R(2 N+\Delta \right)  \left(\omega_a -k+2\pi i T_L(2 N+\Delta \right) \nonumber \\
&\times& \prod_{a < 0,N\geq 0, k} \left( \omega_a -k-2\pi i T_L(2 N+\Delta\right) \left(\omega_a +k-2\pi i T_R(2 N+\Delta \right)\nonumber\\
&\times& \prod_{N \geq 0, k} \left(\omega_0 +k+2\pi i T_R(2 N+\Delta \right)\left(\omega_0 -k+2\pi i T_L(2 N+\Delta \right).
\eeqa
Since we want the determinant for a real scalar, hence the square on the left hand side of the above equation. The first line corresponds to the ingoing modes with $a > 0$, the second line are the outgoing modes and thermal frequencies
with $a < 0$, and the last line corresponds to the zero modes with $a= 0$. 

We can simplify this to write
\beqa
\left(\f{Z^{(1)}(\Delta)}{e^{\text{Pol}(\Delta)}}\right)^{-2}&=&\prod_{a > 0,N\geq 0,k} \left(\left( N+\f{\Delta}{2}+\f{T_L a}{T_L+T_R} \right)^2+\left(\f{k}{2\pi\left(T_L+T_R\right)} \right)^2\right)\nonumber\\
&\times& \prod_{a < 0,N\geq 0, k} \left(\left( N+\f{\Delta}{2}+\f{T_R a}{T_L+T_R} \right)^2+\left(\f{k}{2\pi\left(T_L+T_R\right)} \right)^2\right)\nonumber\\
&\times& \prod_{N \geq 0, k} \left(\left( N+\f{\Delta}{2} \right)^2+\left(\f{k}{2\pi\left(T_L+T_R\right)} \right)^2\right).
\eeqa
We next regulate the product over $k$ as
\beqa
\prod_{a}\left(1+\f{x^2}{k^2}\right)&=&\frac{\sinh \pi x}{\pi x}=\f{e^{\pi x}}{\pi x}\left(1-e^{-2 \pi x}\right).
\eeqa
Next absorbing the polynomial and $\Delta$ independent terms into $e^{\text{Pol}(\Delta)}$ we get
\beqa
\left(\f{Z^{(1)}(\Delta)}{e^{\text{Pol}(\Delta)}}\right)^{-1}&=&\prod_{a > 0,N\geq 0}\left(1-q^{a+N}\bar{q}^{p}\left( q \bar{q} \right)^{\Delta/2} \right)\nonumber\\
&\times& \prod_{a > 0,N\geq 0}\left(1-\bar{q}^{a+N} q^{p}\left( q \bar{q} \right)^{\Delta/2} \right)\prod_{N\geq 0}\left(1-\left( q \bar{q} \right)^{p+\Delta/2} \right)
\eeqa
where
\beqa
q=e^{-2 \pi\left(2 \pi T_L\right)} ~~~ \text{and} ~~~ \overline{q}=e^{-2 \pi\left(2 \pi T_R\right)}.
\eeqa
We can write this in a more compact way as
\beqa
Z^{(1)}=e^{\text{Pol}(\Delta)}\prod_{l,l'= 0}\f{1}{\left(1-q^{l+\Delta/2}\bar{q}^{l'+\Delta/2} \right)}.
\eeqa

\vspace{1cm}

\section{Scalar field in dS}\label{dsqnm}
\subsection{Scalar field solutions}
We now study the behavior and quasi-normal modes of scalar fields in $d$ dimensional de Sitter spacetime following \cite{Du:2004jt, Lopez-Ortega:2006aal}. For this we use the following metric of the de Sitter background in static coordinates
\beqa
ds^2=\left(1-\f{r^2}{\ell^2}\right)dt^2-\f{dr^2}{\left(1-\f{r^2}{\ell^2}\right)}-r^2d\Omega^2
\eeqa
where $d\Omega^2$ is the line element of the $(d-2)$ dimensional unit-sphere and the cosmological horizon occurs at $r = \ell$.
We again solve the Klein-Gordon equation in this geometry and the solution in this case takes the form
\beqa
\phi=e^{-i\omega t}Y_{l,m}R(r).
\eeqa
Where, $Y_{l,m}$, again, stands for the spherical harmonics on the $(d-2)$ dimensional unit sphere and $R(r)$ the radial part of the solution. After necessary simplifications, $R(r)$ satisfies a hypergeometric type differential equation and thus again has two solutions. We define the following parameters
\beqa
a&=&A+B+\f{d-1}{4}+\f{1}{2}\left(\f{(d-1)^2}{4}-m^2\ell^2\right)^{\f{1}{2}}\nonumber\\
b&=&A+B+\f{d-1}{4}-\f{1}{2}\left(\f{(d-1)^2}{4}-m^2\ell^2\right)^{\f{1}{2}}\nonumber\\
c&=&2A+\f{d-1}{2}\nonumber\\
A&=&\f{l}{2}\nonumber\\
B&=&\omega \ell.
\eeqa
With these, the first solution, with $y=r^2/\ell$ is
\beqa
R_1&=& y^{l/2} (1-y)^{\omega \ell} \ _2F_1(a,b,c;y).
\eeqa
If $c$ is not an integer then the second solution is
\beqa
R_2=y^{l/2} (1-y)^{-\omega \ell}~_2F_1(a-c+1,b-c+1,2-c;y)
\eeqa
while if $c$ is an integer greater than zero
\beqa
R_2&=&y^{l/2} (1-y)^{-\omega \ell} \ _2F_1(a,b,c;y)\ln(y)\nonumber\\
&+& \f{(c-1)!}{\Gamma(a)\Gamma(b)}\sum_{s=1}^{c-1}(-1)^{s-1}(s-1)!\f{\Gamma(a-s)\Gamma(b-s)}{(c-s-1)!}y^{-s}\nonumber\\
&=&\sum_{s=0}^{\infty}\f{(a)_s(b)_s}{s!(c)_s}y^s\left[\psi(a+s)+\psi(b+s)-\psi(c+s)-\psi(1+s)\right.\nonumber\\
&-&\left.\psi(a-c+1)-\psi(b-c+1)+\psi(1)+\psi(c-1)\right].
\eeqa
where $(y)_0=1, (y)_s=y(y+1)\cdots(y+s-1)$ for $s\geq 1$ and $\psi(y)=d \ln \Gamma(y)/dy$.
The de Sitter QNMs are solutions to the equations of motion
that satisfy the physical boundary conditions: (i) the field is regular at the origin; (ii) the
field is purely outgoing near the cosmological horizon.
However, the radial function $R_2$ is divergent at $r=0$. Thus the only surviving solution is
\beqa
R(y)=y^{\f{l}{2}}(1-y)^{\f{i\omega\ell}{2}}~_2F_1(a,b,c;y).
\eeqa
To impose the second condition, we can write this as
\beqa
R&=&y^{\f{l}{2}}\left[(1-y)^{\f{i\omega\ell}{2}}\f{\Gamma(c)\Gamma(c-a-b)}{\Gamma(c-a)\Gamma(c-b)}~_2F_1(a,b,a+b+1-c;1-y)\right.\nonumber\\
&+&\left.(1-y)^{\f{-i\omega\ell}{2}}\f{\Gamma(c)\Gamma(a+b-c)}{\Gamma(a)\Gamma(b)} ~ _2F_1(c-a,c-b,c+1-a-b;1-y)\right].
\eeqa
the first term above represents an
ingoing wave and the second term represents an outgoing wave. Thus we require $c-a =-n$, or $c-b =-n, ~~n = 0, 1, 2,\cdots$ and we may read off the de Sitter quasi-normal frequencies are
\beqa
\omega_{n,l}=-i\f{2n+l+\Delta_{\pm}}{\ell},
\eeqa
where we now represent $\Delta_{\pm}$ as
\beqa
\Delta_{\pm}=\f{d-1}{4}\pm\f{1}{2}\left(\f{(d-1)^2}{4}-m^2\ell^2\right)^{\f{1}{2}}.
\eeqa
\subsection{One-loop partition function}
To find the one-loop partition function, we proceed in the usual way. After Wick rotation, the dS temperature can be determined as $T=\f{1}{2\pi \ell}$, by computing the radius of Euclidean time circle required for regularity. We may again use \ref{adspf}
 and the degeneracies $D_l$ to write down the partition function as
 \beqa
 \log Z=\text{Pol}(\Delta_{\pm})-\sum_{\pm}\sum_{l=0}^{\infty}D_l\log(l+\Delta_{\pm}).
 \eeqa

\end{appendix}

\bibliographystyle{JHEP}
\bibliography{fscqnm}

\end{document}